# Tailoring iridescent visual appearance with disordered resonant metasurfaces

Adrian Agreda[1§], Tong Wu[1§], Adrian Hereu[2], Mona Treguer-Delapierre[2], Glenna L. Drisko[2], Kevin Vynck[3], Philippe Lalanne[1]*

[1]LP2N, CNRS, Institut d'Optique Graduate School, Univ. Bordeaux, F-33400 Talence, France
[2]CNRS, Univ. Bordeaux, Bordeaux INP, ICMCB, UMR 5026, F-33600 Pessac, France
[3]Institut Lumière Matière, CNRS, Université Claude Bernard Lyon 1, 69100 Villeurbanne, France
[§]These authors contributed equally to the work.

*Corresponding author: Philippe.Lalanne@institutoptique.fr

**Abstract:** The nanostructures of natural species offer beautiful visual appearances with saturated and iridescent colors and the question arises whether we can reproduce or even create new appearances with man-made metasurfaces. However, harnessing the specular and diffuse light scattered by disordered metasurfaces to create attractive and prescribed visual effects is currently inaccessible. Here, we present an interpretive, intuitive and accurate modal-based tool that unveils the main physical mechanisms and features defining the appearance of colloidal disordered monolayers of resonant meta-atoms deposited on a reflective substrate. The model shows that the combination of plasmonic and Fabry-Perot resonances offers uncommon iridescent visual appearances, differing from those classically observed with natural nanostructures or thin-film interferences. We highlight an unusual visual effect exhibiting only two distinct colors and theoretically investigate its origin. The approach can be useful in the design of visual appearance with easy-to-make and universal building blocks having a large resilience to fabrication imperfections, and potential for innovative coatings and fine-art applications.

Nature offers beautiful colored appearances produced by the interaction of light with micro- and nanoscale structures[1]. The most resplendent appearance of butterfly wings or bird plumage often comes from the iridescence of melanosome forms organized into thin layers[2]. The micro- and nanoscale complexity studied by biologists interested in structural color biodiversity, inspires engineers designing metasurfaces with angle-insensitive structural colors or prescribed angular color response[3,4,5,6]. Be they disordered[7,8,9,10,11] as in Fig. 1a-b, or minutely organized,[12,13,14] man-made metasurfaces composed of arrays of Mie or plasmonic nanoresonators offer several degrees of freedom. At the nanoscale, a plethora of high-index materials and shapes can be used to tailor the resonances of the individual constituent meta-atoms. At the wavelength scale, the arsenal of nanoresonances can be enriched by mode hybridization, leading to considerable changes of the spectral and spatial properties of the modes.[15,16,17] At the mesoscale, a combination of short- and long-range electromagnetic interactions[18,19,20,21,22] results in complicated interferences.[23,24] This rich physics absent in natural low-index morphologies must inevitably reflect in the far-field scattering properties. Therefore, one is entitled to wonder whether it may offer visual appearance[25,26] so far unseen in the biological world.

Visual appearance includes not only color attributes (lightness, hue and saturation) but also geometrical attributes such as gloss, texture and shape, which cause perceived light to vary from point to point over a surface of uniform color.[27] All these attributes are in principle objectively quantified into the bidirectional reflection distribution function (BRDF), a multidimensional radiometric function that describes how the metasurface scatters light for all possible planewave illuminations (Fig. 1a). By definition, the BRDF relates the radiance $L_r$ of a surface in a particular scattering direction defined by the wavevector $\mathbf{k}_s$ and a polarization direction $\mathbf{e}_s$ to the irradiance $E_i$ in a particular direction $\mathbf{k}_i$ and polarization $\mathbf{e}_i$, i.e., $f = L_r(\mathbf{k}_s, \mathbf{e}_s, \lambda)/E_i(\mathbf{k}_i, \mathbf{e}_i, \lambda)$ for any wavelength $\lambda$.[28]

Shaping the BRDF with high-index subwavelength structures belongs to a longstanding and fundamental ambition of wave science. The problem comes in different forms, but generally consists of angularly and spectrally controlling polychromatic light scattering with nanostructures smartly arrayed on a surface. For disordered metasurfaces, a few dedicated numerical tools relying on Green-tensor computations exist to simulate the electromagnetic properties of large ($\sim 100 - 1000$) ensembles of complex nanoparticles.[29,30,31,32] These tools have varying degrees of generality but remain often limited by the inherent inaccuracies related to finite-size domain simulations. Additionally, a prohibitive amount of computations is often required to calculate the BRDF, even when nanoparticles are simplistically modeled as pure electric dipoles.[33] In that regard, we highlight the importance of considering higher-order multipoles, without which the unusual iridescent effect reported hereafter could not be predicted.

Retrieving the BRDF of disordered metasurfaces with brute-force full-wave electromagnetic analysis is therefore not viable. Besides being inefficient, the approach also hides the physical mechanisms behind visual appearance. We need a tool not only to calculate, but also to understand the design process. Approximate models are required.

In a recent work, some of the authors designed an advanced simulation tool predicting the visual appearance of disordered metasurfaces.[34,35] The tool combines a semi-analytical model that predicts the metasurface BRDFs and a rendering engine that generates true-to-life images of macroscopic and arbitrary objects coated with a metasurface. The force of the model is to disentangle the respective roles played by nanoscale resonances and mesoscale interferences, thereby providing considerable and important insight into the control of the BRDF. Yet, the model requires repeating many full-wave electromagnetic simulations for all wavelengths, incidence angles and polarizations of the incident light,[34] consequently loosing efficiency and physical intuition on the role of nanoscale resonances.

In this work, we go one step further and remove the deficiency of previous approaches. Our BRDF model stands on recent advances[36] in the analysis of electromagnetic nanoresonators in the basis of their natural resonance modes, also known as quasinormal modes (QNMs). This approach considerably simplifies the BRDF computation and unveils the physical mechanisms impacting the color and visual appearance of complex metasurfaces.

The tool is implemented, tested, and validated for disordered colloidal monolayers of silver nanoparticles deposited on a reflective substrate coated with a submicrometric dielectric spacer (Fig. 1a-b). These metasurfaces rely on a cornerstone geometry of nanophotonics: they are easy to fabricate and have a small geometrical cross section as well as optical resonances passively or dynamically tunable by varying the film thickness or the permittivity.[7,16,37,38,39,40,41] By changing the spacer thickness, large and multiple color variations are obtained. They are a bit reminiscent of classical thin-film iridescence.[42] However, their physical origin is completely different. This kind of iridescence is driven by nanoscale resonances (Fig. 1d), rather than thin-film interferences. Accordingly, the color changes are not only observed in the specular direction but in every direction (Fig. 1c), hence the name diffuse iridescence.[34] In the following, we will see that the plasmonic resonances have manifold interesting properties that require a careful multipolar treatment. The disordered monolayers considered in this work thus represent a serious testbed for the tool.

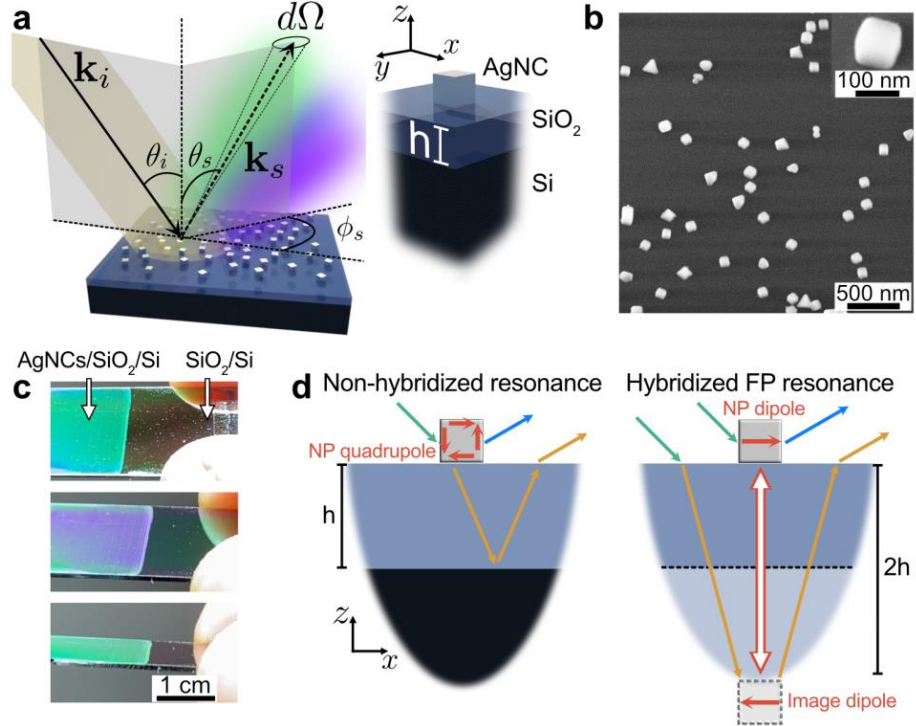

**Figure 1**. **a**, BRDF definition schematic. The samples consist of disordered arrays of ∼ 100 nm side silver nanocubes (AgNCs) on Si substrates, coated with a thin SiO$_2$ layer of height $h$. **b**, Scanning electron microscope (SEM) image of a typical metasurface with a ∼8% surface coverage. A high magnification of a single Ag nanocube is shown in the inset. **c**, Two-color iridescence is observed for $h ≈ 640$ nm under sunlight illumination for various viewing $\theta$ and lighting $\theta_i$ directions. The photographs are extracted from the Supplementary Movie 1. Half of the sample surface is not covered by nanocubes; it does not scatter light and appears dark when the viewing direction differs from the specular reflection direction. **d**, The iridescence is due to two types of resonance modes with markedly different physical properties documented in Supplementary Note 3.

The modal-based BRDF tool quantitatively predicts and intuitively explains an unusual iridescence phenomenon, in which the diffuse light exhibits only two distinct colors, irrespective of the viewing and illumination directions. Such peculiar coloration could be beneficial in security printing, optical filters and innovative coating applications. The modal analysis also explains why, despite the renowned strong sensitivity of the nanoresonator resonance wavelength to shape and size variations, the colors remain saturated and accurately predicted by simulations assuming monodispersity when this is not actually the case (Fig. 1b). This is an important property that considerably lowers fabrication costs of colloidal monolayers and enables large-scale perspective applications.[43]

The tool provides further insight into unexpected properties of the resonance modes of nanoparticles on reflective substrates not reported before.[16,37,38,39,40,41] For instance, it predicts that not all the resonances of individual nanoparticles hybridize (Fig. 1d), despite the presence of the reflective substrate, implying that some resonances have the same frequency irrespective of the spacer thickness. This approach also explains why the frequencies of the BRDF that govern the metasurface colors are not determined by the resonance frequencies.

**RESULTS AND DISCUSSION**

### Two-color diffuse iridescence

The metasurfaces are fabricated by dip-coating a $SiO_2$/Si wafer with a colloidal monolayer of silver nanocubes (see Methods). For the study, we have gathered experimental results for a series of samples obtained with moderate silver nanocube densities ($\rho = 0.2$ to 8 nanoparticles per $\mu m^2$) and $SiO_2$ layer thicknesses, from $h = 0$ to $h = 700$ nm with a $\approx 100$ nm step. High densities reinforce the diffuse brightness, lower the color saturation and whiten the hue. In contrast, low densities reinforce the specular component and weaken the diffuse light resulting in a darker appearance for non-specular angles. Hereafter, we focus on the samples with an intermediate density $\rho = 2\ \mu m^{-2}$. A more detailed analysis of this important parameter is addressed in a future report.

Under a full sunlight illumination, all the samples (except for the reference sample with the nanocubes directly laying on the silicon wafer, $h = 0$ nm) exhibit iridescence with distinctive and vivid colors that vary as the samples are rotated. To quantify and evaluate the color variation, the metasurfaces are characterized in a controlled environment with a collimated solar simulator and the metasurface appearance is recorded with the camera of a smartphone fixed on a motorized stage (see Methods). Figure 2a shows two stripes of photographs recorded for the samples with thicknesses $h = 320$ and $640$ nm. The Supplementary Note 1 further documents the color gamut observed for other thicknesses.

In general, the dependence of the diffuse color with the viewing and lighting directions is gradual and involves several hues, see the upper stripe of photographs in Fig. 2a and Supplementary Fig. S1.1. From the very first observations, we have been intrigued by the metasurface with a spacer thickness $h = 640$ nm. This metasurface stands out strikingly due to its iridescence mostly composed of only two colors, green and violet, the intermediate blue hue appearing only transitorily as we rotate the sample. This uncommon effect is easily seen from the lower series of photographs in Fig. 2a, which is recorded for normal incidence.

The iridescences of both metasurfaces are further documented in Fig. 2b, in which the colors extracted from the photographs recorded by the camera are displayed for six angles of incidence $\theta_i$ and for $|\theta| < 80°$. We note several interesting features for the metasurface with $h = 640$ nm. First, the green and violet colors largely dominate the metasurface response. The blue color is fleetingly observed as a rapid transition between green and violet. Second, the color, initially green around the normal, does not change up to angles of $\theta \approx 45°$, for which the metasurface color abruptly changes to violet, before becoming green again at even larger (almost grazing) viewing angles. Yet, another similar transition occurs for fixed viewing angles, this time as the angle of incidence increases and reaches $\theta_i \approx 45°$. For instance, when it is seen from above around the normal ($\theta = 0$), the metasurface, initially green, suddenly turns violet for $\theta_i \approx 45°$, before becoming again green for large incidences ($\theta_i \approx 70°$).

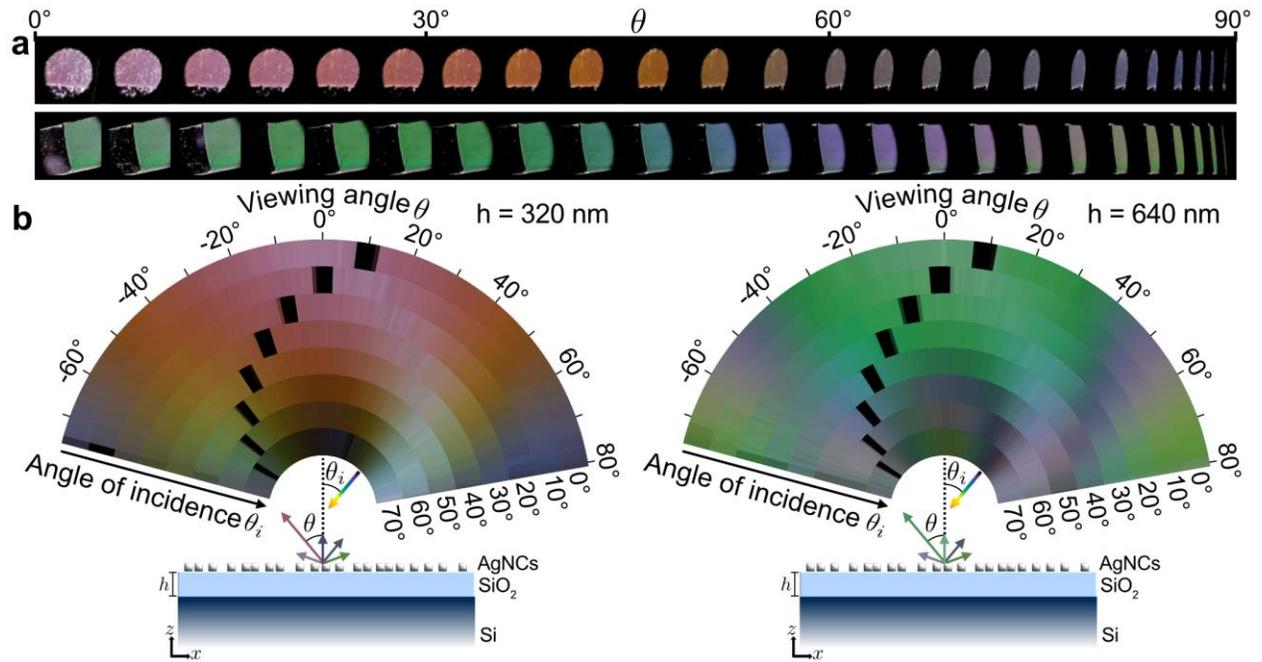

**Figure 2.** Direct visual demonstration of the two-color iridescence phenomenon with a solar simulator. **a**, Series of photographs of two metasurfaces with $h = 320$ nm (top, multicolor iridescence) and $h = 640$ nm (bottom, two-color iridescence) taken by varying the viewing angle at normal incidence. **b**, Extended visualization of the two metasurfaces at several angles of incidence ($\theta_i = 0° \rightarrow 70°$) and viewing angles ($\theta = -75° \rightarrow 80°$). The colors are extracted from the central part of the metasurface in steps of $\Delta\theta = 1°$ and black zones are due to the detector support blocking the incident light. Note that, since the area occupied by the angular sectors for small $\theta_i$'s is greater than that corresponding to large $\theta_i$'s, the weight of the colors at the largest $\theta_i$'s are underrepresented in the hand-fan.

One may be tempted to see these color transitions as classical iridescence resulting from thin-film interference combined with a diffuse reflection added for instance by rough film surfaces. This vision is quite erroneous for several reasons. First, the light diffusion feeds the entire half space above the metasurface, owing to the deep subwavelength scale of the nanocubes, in contrast to the diffusion produced by a rough film that occurs around the specular direction. Secondly, and more importantly, the diffuse colors and their gradual variations have very little to do with those produced by the specular iridescence of the same thin SiO₂ layer. This point is highlighted in Supplementary Note 2 by comparing the diffuse iridescence gamut with the specular iridescence computed and measured on the same sample in a zone without nanoparticles. Thin-film interference is indeed involved in the diffuse iridescence mechanism, but its contribution is weak. The nanoparticle resonances prevail (Supplementary Note 3).

In fact, two-color iridescence exists in nature. For instance, the wing color of the celebrated Morpho butterfly, which is blue when the viewing angle is normal or slightly inclined, and changes to violet rather suddenly when the angle becomes large enough, while maintaining the lighting direction perpendicular to the wing veins.[44] Instead, the present iridescence stems from a large hue difference, $\Delta h_{ab} \approx 150°$ in the HSV/HSL encodings of RGB,[45] between the two colors, green and violet, therein largely neglecting the blue and cyan parts of the spectrum. A large difference is maintained almost irrespectively of lighting directions, in contrast with the continuous color variation produced by film interferences as the incidence or viewing directions vary. More details are found in Supplementary Note 4.

The distinct and chromatically separated hues can produce a beautiful visual effect for curved metasurfaces. Owing to the curvature, the normal to the metasurface is no longer fixed. Therefore, we expect to observe a non-uniform appearance with a two-color patchwork of prevailing violet and green zones that alternate depending on the local normal of the metasurface. As predicted in,[34] variations of the viewing direction will result in a dynamic change of the patchwork, in which the colored zones gently deform and glide on the curved surface (Supplementary Note 5).

**The resonance modes of plasmonic nanoresonators on reflective substrates**

In this Section, we study the resonance modes supported by the individual nanocubes. These resonances play a fundamental role in the modal BRDF tool developed further below.

Figure 3a displays the eigenwavelengths $\tilde{\lambda}$ ($Q = \text{Re}(\tilde{\lambda})/2\text{Im}(\tilde{\lambda})$ is encoded with colors) of a silver nanocube as the spacer thickness is varied. The nanocube is assumed to have a 100 nm size as estimated by averaging particles imaged in several transmission electron microscopy (TEM) images. As we first dedicate the analysis to the iridescence for normal incidence, only the QNMs with a symmetry plane ($Oxz$ or $Oyz$) are considered. The insets display a few QNM radiation diagrams computed at the real QNM resonant frequencies for normalized QNMs. The QNMs are computed and normalized with the PML-normalization method using the QNMEig solver[46] of the MAN (Modal Analysis of Nanoresonators) freeware package[47] based on COMSOL Multiphysics, see Methods.

Two types of resonances with contrasted behaviors are found. The first one presents eigenwavelengths that notably vary across the visible spectral range as $h$ varies. In contrast, and perhaps surprisingly, we also find a second type in the blue part of the spectrum with eigenwavelengths that are nearly independent of $h$. As explained below, the second type of resonance has a frequency (wavelength) fixed by the particle size and shape, while the other corresponds to a 'plasmonic dimer'[38,39,40] formed by the nanocube and its mirror-image in the silicon substrate.

To trace back the origin of the two types of modes, we first consider the QNMs of a silver nanocube laying directly on a $SiO_2$ semi-infinite substrate (we refer to this geometry without silicon as $h = \infty$ hereafter). We find two QNMs, either with dominant dipolar or quadrupolar characters. Their near fields are shown in the maps on the left sides of Fig. 3b and Fig. 3c. Comparisons with the actual QNM field distributions of the silver nanocube on the $SiO_2$/Si substrate are also shown in the same figure for three different values of $h$. Specifically, we find that the first- and second-type resonances result from electric-dipolar (ED) and electric-quadrupolar (EQ) modes.

It is well known that substrate-mediated hybridizations of dipolar and higher-order plasmons of nanoparticles laying on high-index substrates significantly change the resonant frequency and radiative decay of localized plasmons by forming bonding and antibonding states.[15,19,38,39,40] In this case, we may intuitively see the metasurface as a bilayer, in which a dense network of in-layer and inter-layer electromagnetic interactions takes place. When the separation distance between the nanoparticles and their mirror image becomes comparable to the wavelength, the hybridization does not rely on static (evanescent) fields, but rather on propagating waves (right inset in Fig. 1d). The hybridized modes then acquire a strong photonic character with a significant field in the silica layer, see Fig. 3c. They are Fabry-Perot modes[16,40,41] formed by the bouncing of light, back and forth between the silicon substrate and the resonant nanoparticle (or between the nanoparticle and its virtual mirror image in the bilayer picture). All the QNMs of Fig. 3a, which line up in a series of parallel branches as $h$ varies, are Fabry-Perot resonances. The number of nodes inside the thin layer is constant within a branch and varies from branch to branch (Fig. 3c).

It remains to be understood why the frequencies of the second-type resonances do not significantly vary with $h$, but rather oscillate around the frequency of the EQ mode, which is shown with the red vertical line in Fig. 3a. We hypothesize that it is due to the quadrupolar nature of the resonance. In Fig. 3d, we compare the scattering diagrams of the ED and EQ modes for $h = \infty$. In clear contrast with the ED mode, the EQ mode radiates much less energy towards the SiO$_2$ substrate ($z < 0$ half-plane). Additionally, it preferentially radiates light at $\approx \pm 45°$ from the normal in the silica layer, with a very weak radiation in the normal $z$-direction. Though the hybridization relies on the QNM field radiated at wavelength-scale (not far field) distances, we think that the difference in the radiation diagrams intuitively explains why the ED mode strongly hybridizes, while the quadrupolar mode does not form Fabry-Perot resonances with the substrate.

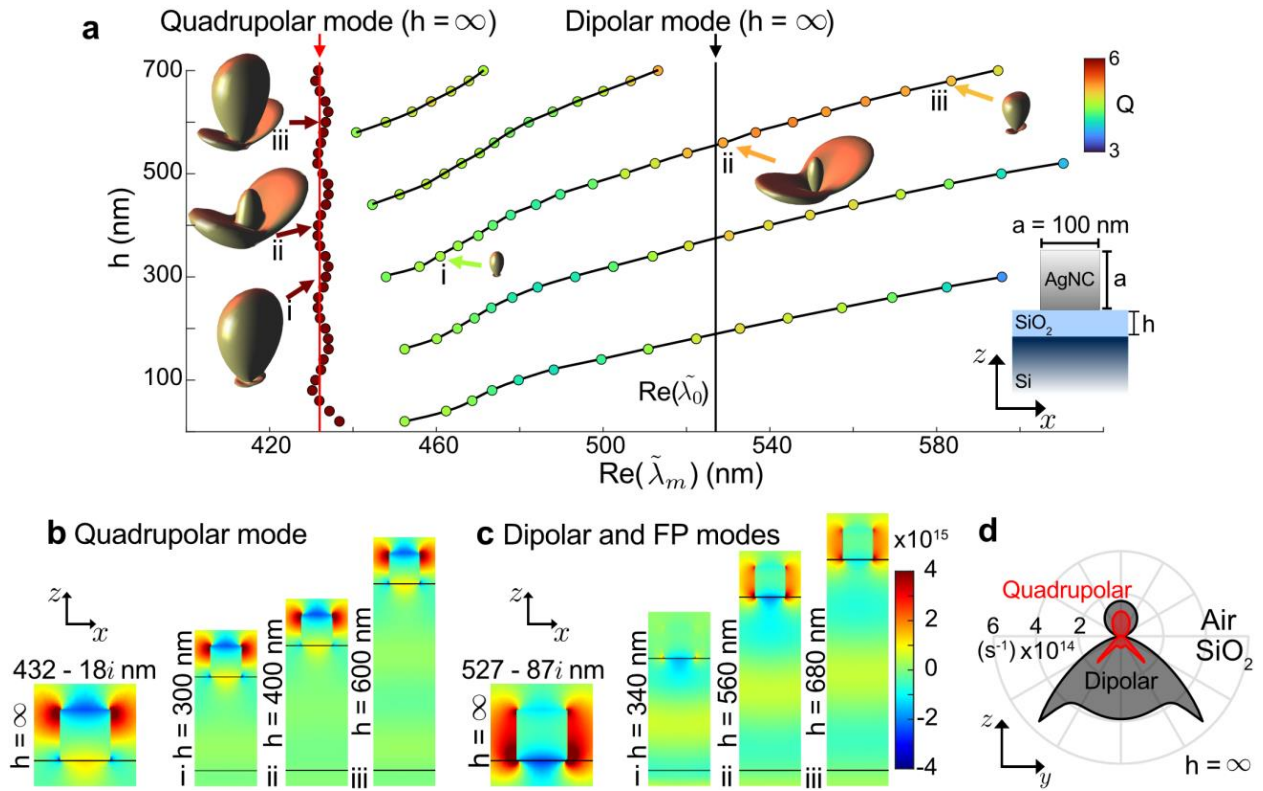

**Figure 3. a**, Eigenwavelengths of the dominant QNMs for a (100×100×100 nm$^3$) Ag nanocube on a Si substrate coated with a silica film of thickness $h$. The color of the dots represents the quality factor $Q$. The scattering diagrams of several normalized QNMs are also displayed in the insets. The solid vertical lines correspond to the real part of the eigenwavelengths of the electric dipolar (ED) (black) and quadrupolar (EQ) (red) modes of the same nanocube on a SiO$_2$ substrate ($h = \infty$). **b-c**, Near-field distributions of the quadrupolar (**b**) and dipolar (**c**) modes for three values of $h$ labeled as i, ii and iii in **a**. The maps show the real part of the QNM electric field $x$-component, $\tilde{E}_x$, in the $xz$-plane. The fields are computed and normalized with the freeware MAN.[47] **d**, Scattering diagrams of the normalized dipolar and quadrupolar QNMs for $h = \infty$.

The six insets in Fig. 3a show the far-field radiation diagrams of a few normalized QNMs, see Methods. The diagrams feature various dominant lobes that drive the change of the metasurface color as the viewing angle $\theta$ varies. Irrespective of whether the QNM is hybridized or localized on the nanocube, the dominant scattering directions of the diagrams change significantly as $h$ varies, even for QNMs belonging to the same Fabry-Perot branch. The precise mechanisms that lead to the formation of the lobes are comprehensively analyzed in the Supplementary Note 3. The latter shows that the dominant scattering directions can be analytically predicted with simple thin-film interference arguments. For the localized EQ resonance, the interference occurs between the light scattered off the Ag nanocube, either directly into air or after reflection on the substrate (Fig. S3.1). In contrast, for the hybridized FP resonance, the dominant scattering directions are determined by constructive interference between the light scattered by the nanocube and its virtual mirror image (Fig. S3.3).

**BRDF measurements**

To better assess the origin of the diffuse iridescence phenomenon, we consider the metasurface BRDF. We illuminate the samples with a supercontinuum laser beam and record the spectrally and angularly resolved response with a gonio-scatterometer (see Methods). The BRDFs are then normalized by the nanocube density, $\rho \approx 2 \; \mu m^{-2}$. Figures 4a-c show the BRDF intensity maps for three angles of incidence. The horizontal black strips correspond to detection angles for which the incident light is blocked by the detector support. The maps carry several noticeable features. First, they are quasi-symmetric with respect to the surface normal, rather than to the specular direction (black strips) as would be the case for the iridescence of a rough thin film, for instance. This is likely due to the deep subwavelength size of the nanocubes that act as Lambertian diffusers. Second, the maps show distinctive peaks that vary in form and number as the incidence or viewing angles vary. Note that, voluntarily, we do not associate the peaks with resonances since this association is unjustified in the present case, as will be clarified in the next section.

For the normal incidence case (Fig. 4a), the BRDF map is composed of an intense peak, at 550 nm for $\theta = 0$, a blue-violet peak at $\lambda \approx 430$ nm, and a series of less intense peaks at large viewing angles, $|\theta| > 50°$. The intense peak is responsible for the green hue observed at small viewing and illumination angles in the rightmost hand fan of Fig. 2b. As the illumination angle increases (see Fig. 4b-c), the main trends are a blue shift of all the peaks and a splitting of the intense and initially green peak into two smaller peaks, symmetrically centered around $\theta = 0°$. These trends are particularly visible in the Supplementary Fig. S6.1a that displays a progressive series of additional BRDF maps measured for other angles of incidence. These trends are comprehensively explained in the Supplementary Note 6 with the help of the form-factor model developed in the next Section. The model unveils the available dials, permitting variation of the hue and controlling the dominant lobes of the scattered light. In brief, the model attributes the blue shift to classical interferences in the $SiO_2$ film, which modulate the intensity of the driving field incident onto the nanocube and shift its maxima. It also attributes the splitting to a peculiar property imposed by the hybridized modes. In fact, the modes are effectively excited whenever the direction $\theta_i$ of the incident plane wave coincides with their preferential and natural scattering direction $\theta$ which corresponds to the lobe maxima in the radiation diagrams of Fig. 3a, for example. This explains why the split angularly increases as $\theta_i$ increases and why split peaks coincide with the horizontal black strips for which $\pm\theta \approx \theta_i$.

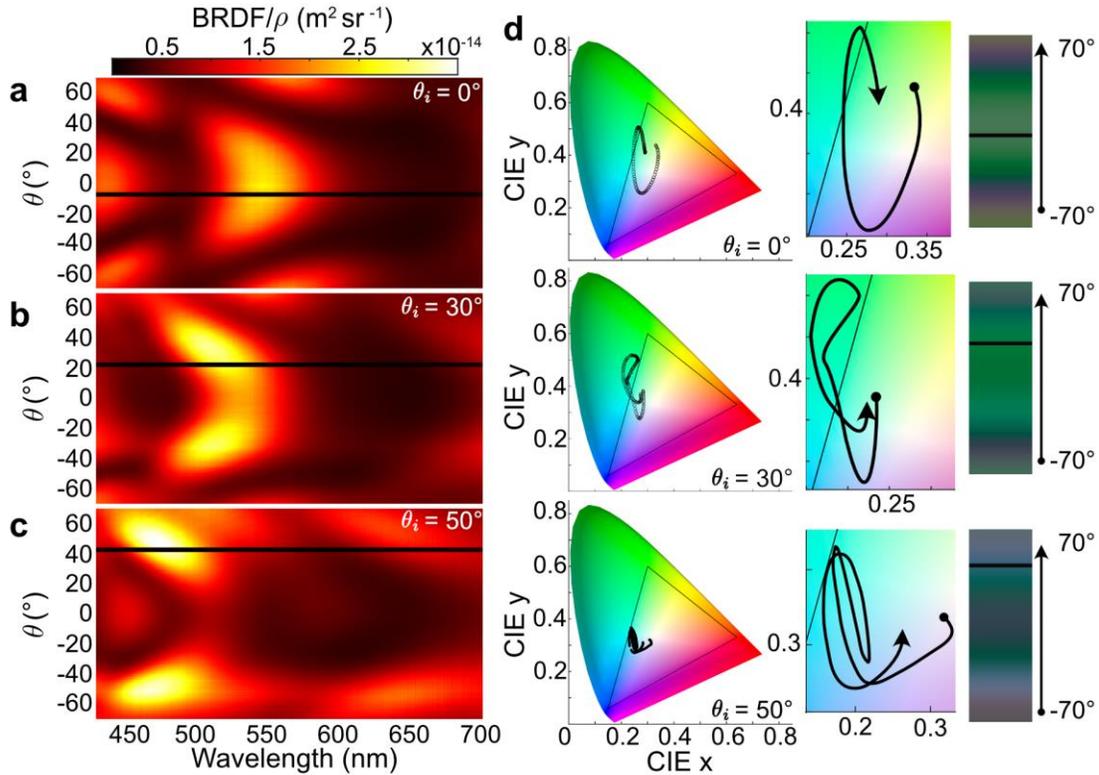

**Figure 4.** Measured BRDFs of the two-color metasurface as a function of the wavelength and viewing polar angle $\theta$ for several incidences: **a**, $\theta_i = 0°$ ; **b**, $\theta_i = 30°$ ; **c**, $\theta_i = 50°$. The measurements are performed $10°$ above the plane of incidence. The BRDFs are normalized by the nanocube density, $\rho = 2\ \mu m^{-2}$. **d**, The chromaticity diagrams show the loci of the colors achieved at three angles of incidence as the viewing angle varies from -70° to 70°. The delimiting triangle indicates the sRGB color space range. The chromaticity coordinates are calculated after rendering the colors from the spectra showed in **a**, **b** and **c**. Magnifications of the color trajectories of the CIE diagrams are shown as insets. Vertical stripes show the corresponding rendered colors and brightness at every measured angle.

The BRDF maps are rather complex. To confirm and quantitatively assess their dominant green and violet hues, we calculate their chromaticity coordinates, $x$ and $y$. These coordinates are widely used to link reflection or transmission spectra with the physiologically perceived colors in human vision. In the chromaticity diagram, the reference white is located at $x = y = 1/3$. The direction and distance from that point express the chroma and saturation of the color, respectively.[44] The curves in the insets of Fig. 4d represent the color trajectories of the metasurfaces as the viewing angle varies from $-70°$ to $70°$ with a $1°$ step, each angle $\theta$ being marked with a point. Magnified images of the central part of the diagrams are shown as insets as well as the rendered colors from the spectra. The trajectories do not exactly represent the colors perceived by the human eye given the limited bandwidth of the incident supercontinuum laser ($\lambda > 430$ nm). Significantly, the visible spectrum starts at $\lambda \approx 380$ nm. Regardless of the missing visible wavelengths, their impact on the perceived colors is negligible as demonstrated by the rendered colors in Fig. 4d.

**Conceptualizing the BRDF with only a few physical quantities**

We have now a clear overview of the resonances at play for all thicknesses and the evolution of the multipeak BRDFs with the incidence angle. The questions arise, how does every individual QNM contribute to the appearance and, in the present context, how is it possible that several resonances (they are four for $h = 640$ nm) result in only two dominant diffuse colors. To answer these questions, we start by recalling the BRDF model in ref. 34 and further make explicit the resonance contribution in the model.

The key assumption in ref. 34 is that multiple scattering gives rise to some moderate corrections to the radiative properties obtained with the single scattering approximation.[48,49] Although its domain of validity is not well known, this approximation commonly referred to as the single scattering approximation in random-media theory[48] is generally accurate for low (sometimes moderate) nanoparticle densities and small incident ($\theta_i$) and scattering ($\theta_s$) angles, for which in-plane electromagnetic interactions are supposedly weak. This assumption disentangles the respective roles played by the nanoscale and mesoscale physics, thereby providing considerable insight into how to control the BRDF. Within this approximation, the diffuse contribution $f_{diff}$ to the BRDF takes the following form[34]

$$f_{diff} \approx \rho \ \frac{d\sigma_s}{d\Omega}(\mathbf{k}_s, \mathbf{e}_s, \mathbf{k}_i, \mathbf{e}_i) \ S_r(\mathbf{k}_s - \mathbf{k}_i) \ \frac{C(\mathbf{k}_s, \mathbf{e}_s, \mathbf{k}_i, \mathbf{e}_i)}{\cos(\theta_i)\cos(\theta_s)}, \tag{1}$$

in which $\rho$ is the nanoparticle density, $\frac{d\sigma_s}{d\Omega}$ is the form factor that accounts for individual local effects, $S_r$ is the static structure factor that accounts for collective mesoscale effects, and $C$ is a heuristic correction factor that accounts for multiple scattering at grazing incidence and large densities.

The static structure factor is analytically known, and the correction factor is easily computed. The bottleneck is the form factor, which requires the knowledge of the radiation diagrams of the individual meta-atom on the stratified substrate for every possible plane-wave illumination ($\theta_i, \lambda, \mathbf{e}_i$). The knowledge of the angular and spectral properties of the form factor is traditionally assessed with many repeated computations performed by varying the frequency, polarization, and direction of the illumination. This is burdensome and uninstructive. Therefore, we propose to reconstruct the form factor as a sum of the far-field radiation diagrams of a few QNMs. Equation (1) then becomes

$$f_{diff} \approx \rho \ \left[\sum_{m=1,2...}\alpha_m \left(\frac{d\sigma_s}{d\Omega}\right)_m\right] \ S_r(\mathbf{k}_s - \mathbf{k}_i) \ \frac{C(\mathbf{k}_s, \mathbf{e}_s, \mathbf{k}_i, \mathbf{e}_i)}{\cos(\theta_i)\cos(\theta_s)}. \tag{2}$$

The force of the factor in brackets in Eq. (2) comes from disentangling the multidimensional dependance of the form factor $\frac{d\sigma_s}{d\Omega}(\mathbf{k}_s, \mathbf{e}_s, \mathbf{k}_i, \mathbf{e}_i)$ as a sum of the factorized product of an expansion coefficient, $\alpha_m(\omega, \mathbf{k}_i, \mathbf{e}_i)$, which solely depends on the incident plane-wave characteristics and an intrinsic radiation diagram, $\left(\frac{d\sigma_s}{d\Omega}\right)_m (\widetilde{\omega}_m, \mathbf{k}_s, \mathbf{e}_s)$, which solely depends on intrinsic modal quantities and the direction and polarization of the scattered wave. We stress that the dependence of the expansion coefficients with the incident plane wave is known analytically (see Eq. (3) and the Methods section), therein reinforcing the intuitive force of the approach.

Theoretically, an infinite number of QNMs should be retained in the expansion.[36] However, only a few modes (four QNMs are sufficient hereafter) are necessary to quantitatively match the experimental data. In our view, it is crucial for inverse design to have the kind of advanced conceptualization provided by the modal expansion in which a highly multidimensional function, the form factor in this case, is represented with only a few physical quantities.

For our metasurfaces with randomly positioned nanocubes, the static structure factor can be well approximated independently of $\mathbf{k}_s$ and $\mathbf{k}_i$ and equal to unity. Moreover, to further simplify, we take $\frac{C(\mathbf{k}_s,\mathbf{e}_s,\mathbf{k}_i,\mathbf{e}_i)}{\cos(\theta_i)\cos(\theta_s)} = 1$, therein neglecting small variations at slightly oblique incidences. The assumption is validated *a posteriori* by comparison with experimental data. Thus, Eq. (2) simplifies and takes a very simple form: $f_{diff} \approx \rho \left[\sum_{m=1,2...} \alpha_m \left(\frac{d\sigma_s}{d\Omega}\right)_m\right]$. Hereafter, we assume unpolarized incident light and polarization-insensitive receptors and all the results, be they experimental or theoretical, are reported by averaging over two orthogonal polarizations.

Figure 5 summarizes our results on a step-by-step reconstruction of $f_{diff}$ gradually considering the contribution of every QNM. We first consider the differential scattering cross-section of a single 100-nm Ag nanocube for $h = 640$ nm, see the upper map of Fig. 5a obtained from a full-wave computation performed with COMSOL. The BRDF experimental data of Fig. 4a are shown again in the lower map of Fig. 5a, for comparison. Note that, due to the $\theta \rightarrow -\theta$ symmetry at normal incidence, only the measurements for $\theta = 0° - 70°$ are shown. The agreement exceeds qualitative commonalities, especially if one considers that the single nanocube computation completely neglects the nanocube polydispersity and the electromagnetic interaction between nanocubes.

Encouraged by the agreement, we further study how the most relevant QNMs contribute to the scattering cross-section map. Figure 5b displays a series of reconstructions of the cross-section maps obtained by progressively increasing the number of QNMs retained in the expansion of the scattered field.[36,47] Additionally, for the sake of clarity, the QNM eigenfrequencies are highlighted with dashed-white vertical lines and black arrows. Only the EQ QNM is considered for the upper map; we indeed obtain an intense violet peak at $2\pi c/\text{Re}(\widetilde{\omega}_{EQ}) = 433$ nm. Adding the Fabry-Perot mode FP$_1$ ($2\pi c/\text{Re}(\widetilde{\omega}_1) = 455$ nm) to the next map lowers the prominence of the violet peak and brings a second weak peak at a much longer wavelength $\approx 560$ nm. As we introduce the two last QNMs, labelled FP$_2$ ($2\pi c/\text{Re}(\widetilde{\omega}_2) = 490$ nm, baby blue) and FP$_3$ ($2\pi c/\text{Re}(\widetilde{\omega}_3) = 560$ nm, green-yellow) in the two lower maps, we end up with a reconstruction that is quite similar to the numerical data of Fig. 5a.

Conventionally, one expects a one-to-one correspondence between resonance frequencies and the peak frequencies. The evolution of the cross-section maps clearly contradicts this well-established intuition, as exemplified by the long-wavelength peak brought by FP$_1$ in the reconstruction, or, even more strikingly, by the absence of any peak at the FP$_2$ resonance frequency.

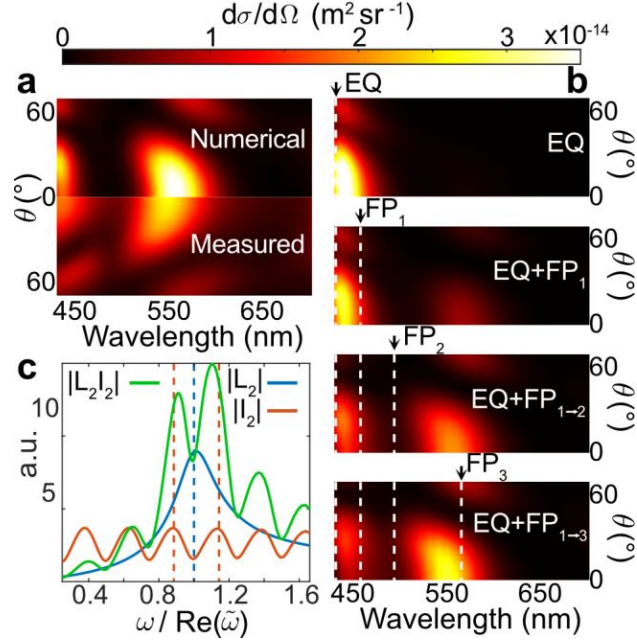

**Figure 5.** Backscattering cross-section spectra as a function of viewing angle $\theta$ for a single nanocube and $h = 640$ nm. **a,** Full-wave numerical results obtained with COMSOL (no free parameter in the model) are compared with BRDF measurements normalized by the nanocube density. **b,** Reconstructed scattering cross-section using 1,2,3 and 4 QNMs. The eigenfrequencies of the EQ, FP$_1$, FP$_2$, and FP$_3$ modes are indicated with black arrows and dashed-white lines. **c,** Plots of a Lorentzian function $L_2(\omega)$ (blue) centered at the resonance frequency $\tilde{\omega}_2$ of the FP$_2$ mode, the volume overlap integral $I_2(\omega)$ (orange) between the driving and the QNM FP$_2$ fields and the absolute value of the modal excitation coefficient $\alpha_2 = L_2 I_2$ (green). The maxima of $L_2$ and $I_2$ are indicated by blue and orange dotted lines. The reconstruction is performed using the reconstruction toolbox of the freeware MAN.[47]

This can be understood by considering the QNM expansion of the field $\mathbf{E}_S(\mathbf{r}, \omega)$ scattered by a resonator upon excitation by a driving field $\mathbf{E}_b(\omega)$, $\mathbf{E}_S(\mathbf{r}, \omega) = \sum_m \alpha_m(\omega) \tilde{\mathbf{E}}_m(\mathbf{r})$,[36] and further focusing on the QNM excitation strength $\alpha_m$. The latter reaches a maximum when the driving and QNM fields well match spatially and spectrally:

$$\alpha_m(\omega) = (\varepsilon_{\text{Ag}} - 1) I_m(\omega) L_m(\omega), \tag{3}$$

with $L_m(\omega) = \frac{\omega}{\tilde{\omega}_m - \omega}$ a Lorentzian function and $I_m(\omega) = \iiint_{V_r} \mathbf{E}_b(\mathbf{r}, \omega) \cdot \tilde{\mathbf{E}}_m d^3\mathbf{r}$ an overlap between $\mathbf{E}_b(\mathbf{r}, \omega)$ and the normalized QNM electric field $\tilde{\mathbf{E}}_m(\mathbf{r})$, which is performed over the volume $V_r$ of the nanocube.[36] Note that, insignificantly for the present purpose, a more accurate formula has been used to compute the $\alpha_m$'s and reconstruct the field in Fig. 5b (see Methods).

Figure 5c illustrates the interplay of $L_m(\omega)$ and $I_m(\omega)$ in the formation of the scattering cross-section peaks for the striking example of the FP$_2$ resonance. For nanoresonators on highly reflective substrates, the driving field intensity is an almost stationary pattern with alternating intensity maxima or minima as the silica thickness or the driving frequency $\omega$ are varied. The sinusoidal pattern distorts the Lorentzian response $L_m(\omega)$. When the maxima of $L_m$ and $I_m$ coincide, the product $L_m(\omega) I_m(\omega)$ results in a reinforcement of the resonance peak. However, it might also happen that the maximum of $L_m(\omega)$ coincides with a minimum of $I_m(\omega)$. Then, the excitation strength exhibits a camel-like response, in which the resonance peak splits in two side peaks (Fig. 5c).

**Resilience to fabrication defects**

Metasurfaces generally face substantial challenges towards large-scale production for the consumer market. Their performance falters due to fabrication imperfections therefore requiring expensive, slow, and size-limited fabrication techniques.[50,51,52] For instance, for color generation with Mie or plasmonic resonances, small size or shape variations of the constituent nanoparticles lead to significant color changes.[2,53] Despite significant size or shape dispersions and occasional particle aggregation (see the statistical analysis in the top inset in Fig. 6) quantitative agreement is achieved between the measurements and our theoretical predictions based on a monodisperse approximation (Figs. 5a and S5.1). This suggests that the present samples are exceptionally resilient to fabrication imperfections. This robustness is applicable to similar iridescence phenomena for nanoparticles composed of other materials.

To intuitively understand the reason for the resilience, we again use QNM theory and study the variations of the BRDF peak frequencies induced by changes of the nanocube size. Both terms, $L_m(\omega)$ and $I_m(\omega)$, must be considered according to the previous Section. The second term brings great stability: its minima or maxima are independent of the nanoparticle size, shape, or even aggregation. It solely depends on the low-index film thickness that is well controlled. The first term, $L_m(\omega) = \frac{\omega}{\widetilde{\omega}_m - \omega}$, is universally encountered in all resonant systems and is a priori not expected to offer a distinctive resilience for the present metasurfaces in comparison to others. In fact, hybridization makes the variations of $\widetilde{\omega}_m$ less sensitive to imperfections. To see this, we consider the dominant ED mode and compare two geometries. Figure 6a is obtained for a reference case (no hybridization), in which the nanocube directly sits on a semi-infinite $SiO_2$ substrate. Consistent with earlier works on plasmonic color generation,[2,4] both the real and imaginary parts of the ED eigenfrequency $\widetilde{\lambda}_{ED}$ largely vary. Notably, $\text{Re}(\widetilde{\lambda}_{ED})$ spans over almost the entire visible spectral domain, $\text{Re}(\Delta\widetilde{\lambda}_{ED}) \approx 200$ nm. Comparatively, much smaller variations are obtained for each Fabry-Perot mode, $FP_1$, $FP_2$ or $FP_3$ of the hybridized case ($h = 640$ nm), the maximal variation, $\text{Re}(\Delta\widetilde{\lambda}_3) = 60$ nm, being obtained for $FP_3$ (Fig. 6b).

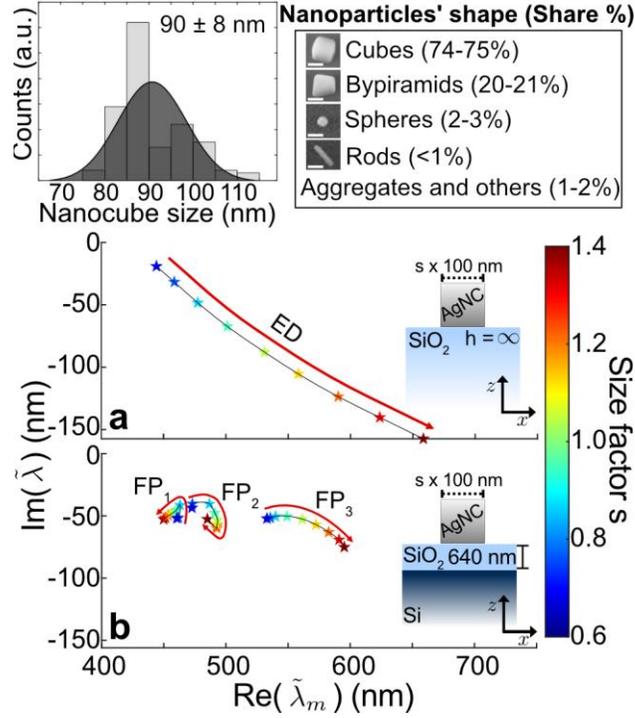

**Figure 6.** Resilience of the hybridized QNMs to size dispersion. **Top inset**, Size and shape distributions of the nanocubes. On the left side, the fitted normal distribution yields an average nanocube edge length of $90 \pm 8$ nm. The table on the right side shows the distribution of the different shapes of particles in the solution. **a**, Trajectory of the computed complex eigenfrequencies for the reference non-hybridized case, $h = \infty$ as the AgNC size ($s \times a$, $a = 100$ nm) is varied. **b**, Same computation for the three FP modes for $h = 640$ nm. In **a-b**, the color of the star marks represents the size factor $s$.

The role of hybridization in the resilience of the complex frequency against size dispersion can be understood with general arguments based on non-Hermitian cavity perturbation theory. The latter stipulates that the size-dependent eigenfrequency change $\Delta\widetilde{\omega}$ is proportional to the square of the normalized QNM electric field $\widetilde{\mathbf{E}}(\mathbf{r})$ inside the perturbation volume $\Delta V$, $\Delta\widetilde{\omega} \propto \iiint_{\Delta V} \widetilde{\mathbf{E}}(\mathbf{r}) \cdot \widetilde{\mathbf{E}}(\mathbf{r}) \, d\mathbf{r}$.[54] For non-hybridized plasmonic nanoparticles, $\widetilde{\mathbf{E}}(\mathbf{r})$ is intensely localized on the particle (Fig. 3b) and $|\Delta\widetilde{\omega}|$ is large, as we all experience with plasmonic sensors. In contrast, FP modes bear both a plasmonic and a photonic character with a normalized field that is distributed not only in the vicinity of the Ag nanocube but also inside the SiO$_2$ layer (Fig. 3c). This is a general property of hybridized plasmons, which inevitably results in a significant increase of the mode volume and a decrease in the sensitivity to imperfections.

## CONCLUSION

We have provided a comprehensive analysis of the visual appearance of disordered monolayers of resonant nanoparticles laying on a reflective substrate coated with a low-index spacer. These metasurfaces constitute a generic structure of nanophotonics with various intrinsic advantages and applications[39,40,41] and offer a large set of available degrees of freedom for tailoring appearance.[34] Owing to the strong confinements provided by high-index nanoparticles, the appearance of the metasurfaces differs from those encountered in nature with low-index nanostructured materials. We have identified a specific geometry that displays an unusual angular color-dependence, in which only two distinct colors are observed for all directions.

An important and fundamental outcome of the work is the conceptually advanced BRDF model, which represents a highly multidimensional function, the BRDF form factor, with only a few physical quantities. This conceptualization provides a strong reduction in the number of degrees of freedom and offers intuition into the key factors impacting the color and direction of the scattered light. The model is not merely a computational instrument but provides important insight for the understanding and the creation of visual appearances with nanoscale resonances. We emphasize the quantitative agreement between the model predictions and the measurements (Fig. 5), which represents a strong validation of a complicated case based on several subtle resonances combining photonic and plasmonic character.

Notably, we have shown that the metasurfaces possess both hybridized Fabry-Perot-like and non-hybridized resonances. Furthermore, the resonances do not necessarily result in a BRDF peak at the resonance frequency, implying that several resonances with distinct frequencies may be combined to produce the same color. We expect this understanding to inspire and motivate further metasurface designs with prescribed diffuse iridescence that are very difficult to intuitively predict with brute-force computations.

The BRDF model also explains the unexpected resilience of the diffuse iridescent effect to size and shape polydispersity. The two-color iridescence is faithfully reproduced despite neglecting the substantial polydispersity resulting from the bottom-up self-assembly fabrication method. Fabrication flexibility is essential for real applications, as it may result in a strong reduction of manufacturing costs for large-scale coatings production. This reinforces our expectation that the present metasurfaces can find genuine application in security holograms or various stunning coatings for luxury goods, potentially with multilayer arrangements of high refractive index film spacers or correlated disorder to further harness the main BRDF attributes.

## METHODS

**Fabrication.** Samples were fabricated using a multi-step bottom-up approach that involves colloidal synthesis, metal oxide thin film formation and nanoparticles deposition. First, silver nanocubes, 100 nm in edge length, were synthesized via a seed mediated growth protocol.[55] The nanocubes have a PVP layer that provides long-term protection against oxidation. In a second step, a dielectric thin-film of controlled thickness was produced through a sol-gel reaction and deposited on a silicon wafer using the dip-coating technique.[56] A ($h = 100 \pm 10$) nm SiO$_2$ layer is reproducibly deposited. By repeating this process, a series of SiO$_2$/Si substrates is fabricated in the range of $h \approx 105 - 710$ nm. Finally, the silver nanocubes were deposited by dip-coating. The nanocube density was controlled by changing the concentration of the cubes in suspension, changing the dispersive medium, or by increasing the number of deposition cycles by dip-coating.[57]

**Characterization.** All the measurements were performed with an in-house goniospectrophotometric setup, which used a supercontinuum source or a solar simulator for the illumination and a cell-phone camera or a spectrometer for the detection. Two concentric stepper motor rotation stages (Newport, URS75 and URS150) and a vertical arm controlled the incident and scattering (viewing) angles, $\theta_i$ and $(\theta, \phi)$, respectively.

More specifically, for the BRDF measurements, we use an unpolarized centimeter-scale expanded supercontinuum laser beam (Leukos, Rock 400). A short-pass filter (Schott, KG-1) restricted the incident spectrum to the visible range. The backscattered light was collected slightly above the plane of incidence by a 1 mm-diameter optical fiber connected to a spectrometer (Ocean Insight, HDX). The incident laser radiant flux was measured with the same setup and with the fiber detector facing the focused laser beam. In the Supplementary Note 7, the accuracy of the BRDF measurements was assessed with a diffuse reflectance reference sample with a known BRDF.

For the characterization of the color variation with the viewing angle in Fig. 2a, the metasurfaces were illuminated with a directional white light illumination provided by a collimated solar simulator (ASAHI SPECTRA, HAL-320) and the metasurface appearance was recorded with a smartphone camera (iPhone 11, 12 MP 26 mm Wide camera, $f$/1.8 aperture) at fixed exposure times for each series. The camera had an entrance pupil of 2.36 mm. The photographs were taken 12 cm away from the samples to ensure a good resolution of the metasurfaces. These settings gave a N.A. = 0.0098 and were comparable to those of the bare eye under similar conditions. Images were captured with the default automatic white balance of the phone, which accurately matched pure white when the illumination conditions were limited to one source.

The chromaticity coordinates and the color renderings were computed with the CIE 1931 2° standard observer and the illuminant D65, which corresponds to midday light (direct sunlight and scattered light by a clear sky)

**QNM computation.** To compute the resonances of the Ag nanocubes, we used the QNMEig solver[46] of the freeware MAN (Modal Analysis of Nanoresonators).[47] The computation precisely considered the spectral dispersion of silver with a bi-pole Drude model and silicon with four pairs of Lorentz poles.[58] The refractive index of the silica layer is assumed to be equal to 1.47. The COMSOL model can be downloaded from the MANMODELS folder.

The radiation diagrams of Fig. 3 were obtained from the near-field maps of the normalized QNMs and were computed at the real QNM resonant frequencies, $\text{Re}(\widetilde{\omega})$, using the near-to-far-field transform freeware RETOP that accurately handled the presence of the layered substrate.[59] The QNM reconstruction of the BRDF (Fig. 5) was obtained with the reconstruction toolbox of MAN by using the well-tested expressions of the excitation coefficient $\alpha(\omega)$, provided in Eq. 6 in ref. [46], referred to as method M1 in Table 3 in ref. 47.

**ASSOCIATED CONTENT**

Supporting information is appended below.

Movies showing the color change at different incidences and views of a diffuse iridescent metasurface under sunlight illumination on a clear day are available free of charge at https://pubs.acs.org/doi/XX.XXXX/acsnano.XXXXXXX. Half of a SiO$_2$/Si substrate is covered with randomly disordered silver nanocubes. The SiO2 layer is $h \approx 600$ nm thick and the density of nanocubes is $\rho \approx 8 \ \mu m^{-2}$ (.MP4).

**ACKNOWLEDGEMENTS**

K.V. and P.L. acknowledge fruitful discussions with Romain Pacanowski and Pascal Barla (INRIA, Talence) and Xavier Granier and Bertrand Simon (LP2N, Talence, France). P.L. thanks Louise-Eugénie Bataille, Philippe Teulat and Louis Bellando for their help in developing the goniospectrometer setup. P.L. and A.A. acknowledge Jacques Leng (LOF, Pessac, France) for giving free access to the solar simulator. This work has received financial support from the French State and the Région Nouvelle-Aquitaine under the CPER project "CANERIIP", from CNRS through the MITI interdisciplinary programs, and from the French National Agency for Research (ANR) under the project "NANO-APPEARANCE" (ANR-19-CE09-0014). We also acknowledge the financial support from the Grand Research Program « LIGHT » Idex University of Bordeaux, and the Graduate program « EUR Light S&T » PIA3 ANR-17-EURE-0027. David Montero performed FEG-SEM observations at the Institut des Matériaux de Paris Centre (IMPC FR2482), which was co-funded by Sorbonne Université, CNRS and by the C'Nano projects of the Région Ile-de-France.

**REFERENCES**

1. Kinoshita, S.; Yoshioka, S; Miyazaki, J. Physics of Structural Colors. *Rep. Prog. Phys.* **2008**, 71, 076401.
2. Nordén, K. K.; Eliason, C. M.; Stoddard, M. C. Evolution of Brilliant Iridescent Feather Nanostructures. *Elife*, **2021**, 10, e71179.
3. Kristensen, A.; Yang, J. K.; Bozhevolnyi, S. I.; Link, S.; Nordlander, P.; Halas, N. J.; Mortensen, N. A. Plasmonic Colour Generation. *Nat. Rev. Mater*. **2016**, 2, 1-14.
4. Rezaei, S.D.; Dong, Z.; Chan, J.Y.E.; Trisno, J.; Ng, R.J.H.; Ruan, Q.; Qiu, C.W.; Mortensen, N.A; Yang, J.K. Nanophotonic Structural Colors. *ACS Photonics* **2021**, 8, 18–33.
5. Song, M.; Feng, L.; Huo, P.; Liu, M.; Huang, C.; Yan, F.; Lu, Y.Q.; Xu, T. Versatile Full-Color Nanopainting Enabled by A Pixelated Plasmonic Metasurface. *Nat. Nanotechnol.* **2023**, 18, 71-78.
6. Mao, P.; Liu, C.; Niu, Y.; Qin, Y.; Song, F.; Han, M.; Palmer, R. E.; Maier, S. A.; Zhang, S. Disorder-Induced Material-Insensitive Optical Response in Plasmonic Nanostructures: Vibrant Structural Colors from Noble Metals. *Adv. Mater.* **2021**, 33, 2007623.
7. Moreau, A.; Ciracì, C.; Mock, J.J.; Hill, R.T.; Wang, Q.; Wiley, B.J.; Chilkoti, A; Smith, D.R. Controlled-Reflectance Surfaces with Film-Coupled Colloidal Nanoantennas. *Nature.* **2012**, 492, 86-89.
8. Rahimzadegan, A.; Arslan, D.; Suryadharma, R.N.S.; Fasold, S.; Falkner, M.; Pertsch, T.; Staude, I.; Rockstuhl, C. Disorder-Induced Phase Transitions in The Transmission of Dielectric Metasurfaces. *Phys. Rev. Lett.* **2019**, 122, 015702.
9. Jang, M.; Horie, Y.; Shibukawa, A.; Brake, J.; Liu, Y.; Kamali, S.M.; Arbabi, A.; Ruan, H.; Faraon, A.; Yang, C. Wavefront Shaping with Disorder-Engineered Metasurfaces. *Nat. Photon.* **2018**, 12, 84–90.
10. Hsu, C.W.; Zhen, B.; Qiu, W.; Shapira, O.; DeLacy, B.G.; Joannopoulos J.D.; Soljačić, M. Transparent Displays Enabled by Resonant Nanoparticle Scattering. *Nat. Commun*. **2014**, 5, 3152.
11. Fusella, M.A.; Saramak, R.; Bushati, R.; Menon, V.M.; Weaver, M.S.; Thompson, N.J.; Brown, J.J. Plasmonic Enhancement of Stability and Brightness In Organic Light-Emitting Devices. *Nature* **2020**, 585, 379–382.
12. Qiu, C.; Zhang, T.; Hu, G.; Kivshar, Y. Quo Vadis, Metasurfaces? *Nano Lett.* **2021**, 21, 5461–5474.
13. Neshev, D.; Aharonovich, I. Optical Metasurfaces: New Generation Building Blocks for Multi-Functional Optics. *Light. Sci. Appl.* **2018**, 7, 1-5.
14. Kuznetsov, A. I.; Miroshnichenko, A. E.; Brongersma, M. L.; Kivshar, Y. S.; Luk'yanchuk B. Optically Resonant Dielectric Nanostructures. *Science.* **2016**, 354, aag2472.
15. Yamaguchi, T.; Yoshida, S.; Kinbara, A. Optical Effect of The Substrate on The Anomalous Absorption Of Aggregated Silver Films. *Thin Solid Films* **1974**, 21, 173-187.
16. Holsteen, A. L.; Raza, S.; Fan, P.; Kik, P. G.; Brongersma, M. L. Purcell Effect for Active Tuning of Light Scattering from Semiconductor Optical Antennas. *Science* **2017**, 358, 1407-1410.

17. Li, G.; Zhang, Q.; Maier S.; Lei, D. Plasmonic Particle-On-Film Nanocavities: A Versatile Platform for Plasmon-Enhanced Spectroscopy and Photochemistry, *Nanophotonics* **2018**, 7, 1865-1889.
18. Haynes, C.L.; McFarland, A.D.; Zhao, L.; Van Duyne, R.P.; Schatz, G.C.; Gunnarsson, L.; Prikulis, J.; Kasemo, B.; Käll, M. Nanoparticle Optics: The Importance of Radiative Dipole Coupling in Two-Dimensional Nanoparticle Arrays. *J. Phys. Chem. B* **2003**, 107, 7337-7342.
19. Nordlander, P., C.; Oubre, E.; Prodan, K. Li; Stockman, M. I. Plasmon Hybridization in Nanoparticle Dimers. *Nano Lett*. **2004**, 4, 899-903.
20. Jain, P. K.; Eustis, S.; El-Sayed, M. A. Plasmon Coupling in Nanorod Assemblies: Optical Absorption, Discrete Dipole Approximation Simulation, And Exciton-Coupling Model. *J. Phys. Chem. B* **2006**, 110, 18243-18253.
21. Dahmen, C.; Schmidt, B.; von Plessen, G. Radiation Damping in Metal Nanoparticle Pairs. *Nano Lett*. **2007**, 7, 318-322.
22. Auguié, B.; Barnes, W. L. Collective Resonances in Gold Nanoparticle Arrays. *Phys. Rev. Lett.* **2008**, 101, 143902.
23. Cai, H.; Srinivasan, S.; Czaplewski, D.A.; Martinson, A.B.; Gosztola, D.J.; Stan, L.; Loeffler, T.; Sankaranarayanan, S.K.; López, D. Inverse Design of Metasurfaces with Non-Local Interactions. *npj Comput. Mater.* **2020**, 6, 1-8.
24. Gigli, C.; Li, Q.; Chavel, P.; Leo, G.; Brongersma, M.L.; Lalanne, P. Fundamental Limitations of Huygens' Metasurfaces for Optical Beam Shaping. *Laser Photonics Rev.* **2021**, 15, 2000448.
25. Wang, C.; Lin, X.; Schäfer, C. G.; Hirsemann, S.; Ge, J. Spray Synthesis of Photonic Crystal Based Automotive Coatings with Bright and Angular-Dependent Structural Colors. *Adv. Funct. Mater.* **2021**, 31, 2008601.
26. Auzinger, T.; Heidrich, W.; Bickel, B. Computational Design of Nanostructural Color for Additive Manufacturing. *ACM Trans. Graph* **2018**, 37, 1-16.
27. Huang, M.; Pan, J.; Wang, Y.; Li, Y.; Hu, X.; Li, X.; Xiang, D.; Hemingray, C.; Xiao, K. Influences of Shape, Size, And Gloss on The Perceived Color Difference Of 3d Printed Objects. *J. Opt. Soc. Am. A* **2022**, 39, 916-926.
28. Nicodemus, F. E. Directional Reflectance and Emissivity of An Opaque Surface. *Appl. Optics* **1965**, 4, 767-775.
29. Solis, D. M.; Taboada, J. M.; Obelleiro, F.; Liz-Marzán, L. M.; Garcia de Abajo, F. J. Toward ultimate nanoplasmonics modeling, *ACS Nano* **2014**, 8, 7559-70.
30. Bertrand, M.; Devilez, A.; Hugonin, J. P.; Lalanne, P.; Vynck, K. Global polarizability matrix method for efficient modelling of light scattering by dense ensembles of non-spherical particles in stratified media. *J. Opt. Soc. Am. A* **2020**, 37, 70-83.
31. Czajkowski, K. M.; Antosiewicz, T. J.; Effective dipolar polarizability of amorphous arrays of size-dispersed nanoparticles. *Opt. Lett.* **2020**, 45, 3220.
32. Theobald, D.; Beutel, D.; Borgmann, L.; Mescher, H.; Gomard, G.; Rockstuhl, C.; Lemmer, U. Simulation of light scattering in large, disordered nanostructures using a periodic T-matrix method. *J. Quant. Spectrosc. Radiat. Transf.* **2021**, 272, 107802.
33. Herkert, E.; Sterl, F.; Both, S.; Tikhodeev, S.; Weiss, T.; Giessen, H. The influence of structural disorder on plasmonic metasurfaces and their colors – a coupled point dipole approach. Accepted in *J. Opt. Soc. Am. B.* DOI 10.1364/JOSAB.477169.
34. Vynck, K.; Pacanowski, R.; Agreda, A.; Dufay, A.; Granier, X.; Lalanne, P. The Visual Appearances of Disordered Optical Metasurfaces. *Nat. Mater.* **2022**, 21, 1035–1041.
35. Scheffold, F., Metasurfaces Provide the Extra Bling. *Nat. Mater.* **2022**, 21, 994–995.
36. Sauvan, C.; Wu T., Zarouf, R.; Muljarov, E.A.; Lalanne, P. Normalization, orthogonality, and completeness of quasinormal modes of open systems: the case of electromagnetism. *Opt. Express* **2022**, 30, 6846-6885.

37. Pors, A.; Bozhevolnyi, S. I. Plasmonic Metasurfaces for Efficient Phase Control in Reflection. *Opt. Express*. **2013**, 21, 27438-27451.
38. Park, J.; Kang, J.H.; Kim, S. J.; Liu, X.; Brongersma, M. L. Dynamic Reflection Phase and Polarization Control in Metasurfaces. *Nano Lett*. **2017**, 17, 407-413.
39. Mock, J.J.; Hill, R.T.; Degiron, A.; Zauscher, S.; Chilkoti, A.; Smith, D.R. Distance-Dependent Plasmon Resonant Coupling Between a Gold Nanoparticle and Gold Film. *Nano Lett*. **2008**, 8, 2245-2252.
40. Taubert, R.; Ameling, R.; Weiss, T.; Christ, A.; Giessen, H. From Near-Field to Far-Field Coupling in the Third Dimension: Retarded Interaction of Particle Plasmons. *Nano Lett*. **2011**, 11, 4421-4424.
41. Wirth, J.; Garwe, F.; Bergmann, J.; Paa, W.; Csaki, A.; Stranik, O.; Fritzsche, W. Tuning of Spectral And Angular Distribution Of Scattering From Single Gold Nanoparticles By Subwavelength Interference Layers. *Nano Lett*. **2014**, 14, 570-577.
42. Qin, M.; Sun, M.; Bai, R.; Mao, Y.; Qian, X.; Sikka, D.; Zhao, Y.; Qi, H. J.; Suo, Z.; He, X. Bioinspired Hydrogel Interferometer for Adaptive Coloration and Chemical Sensing. *Adv. Mater.* **2018**, 30, 1800468.
43. Fredriksson, H.; Alaverdyan, Y.; Dmitriev, A.; Langhammer, C.; Sutherland, D.S.; Zäch, M.; Kasemo, B. Hole–mask colloidal lithography. *Adv. Mater.* **2007**, 19, 4297-4302.
44. Srinivasarao, M. Nano-Optics in The Biological World: Beetles, Butterflies, Birds, And Moths. *Chem. Rev.* **1999**, 99, 1935-1962.
45. Smith, A. R. Color Gamut Transform Pairs. *Comput. Graph*. **1978**, 12, 12–19.
46. Yan, W.; Faggiani, R.; Lalanne, P. Rigorous Modal Analysis of Plasmonic Nanoresonators. *Phys. Rev. B* **2018**, 97, 205422.
47. Wu, T.; Arrivault, D.; Yan, W.; Lalanne, P. Modal Analysis of Electromagnetic Resonators: User Guide for The Man Program. *Comput. Phys. Commun.* **2023**, 284, 108267. The freeware MAN (Modal Analysis of Nanoresonators) is available Zenodo.
48. Tsang, L.; Kong, J. A. *Scattering of Electromagnetic Waves: Advanced Topics*; John Wiley & Sons: New York, 2004; vol. 26.
49. Wang, B. X.; Zhao, C. Y. The Dependent Scattering Effect on Radiative Properties of Micro/Nanoscale Discrete Disordered Media. *Annu. Rev. Heat Transf.* **2020**, 23, 231-353.
50. Kildishev, A. V.; Boltasseva, A.; Shalaev, V. M. Planar Photonics with Metasurfaces. *Science* **2013**, 339, 1232009.
51. Lee, T.; Lee, C.; Oh, D.K.; Badloe, T.; Ok, J.G.; Rho, J. Scalable and High-Throughput Top-Down Manufacturing of Optical Metasurfaces. *Sensors* **2020**, 20, 4108.
52. Choudhury, S.M.; Wang, D.; Chaudhuri, K.; DeVault, C.; Kildishev, A.V.; Boltasseva, A.; Shalaev, V.M. Material Platforms for Optical Metasurfaces. *Nanophotonics* **2018**, 7, 959-987.
53. Proust, J.; Bedu, F.; Gallas, B.; Ozerov, I.; Bonod, N. All-Dielectric Colored Metasurfaces with Silicon Mie Resonators. *ACS nano* **2016**, 10, 7761-7767.
54. Yang, J.; Giessen, H.; Lalanne, P. Simple Analytical Expression for The Peak-Frequency Shifts of Plasmonic Resonances for Sensing. *Nano Lett*. **2015**, 15, 3439.
55. Lin, Z. W.; Tsao, Y. C.; Yang, M. Y.; Huang, M. H. Seed-Mediated Growth of Silver Nanocubes in Aqueous Solution with Tunable Size and Their Conversion to Au Nanocages with Efficient Photothermal Property. *Chem. - Eur. J.,* **2016**, 22, 2326-2332.
56. Scriven, L. E. Physics and Applications of Dip Coating and Spin Coating. *MRS Proceedings* **1988**, 121, 717-729.
57. Roach, L.; Hereu, A.; Lalanne, P; Duguet, E.; Tréguer-Delapierre, M.; Vynck, K.; Drisko, G. L. Controlling disorder in self-assembled colloidal monolayers via evaporative processes. *Nanoscale*, **2022**, *14*, 3324-3345.
58. Garcia-Vergara, M.; Demésy, G.; Zolla, F. Extracting an Accurate Model for Permittivity From Experimental Data: Hunting Complex Poles From The Real Line. *Opt. Lett.* **2017**, 42, 1145–1148. The material dispersion parameters of Si and Ag can be found in the COMSOL model 'QNMEig_NanolettSi.mph' of MAN ref. [47].
59. Yang, J.; Hugonin, J. P.; Lalanne, P. Near-To-Far Field Transformations for Radiative And Guided Waves. *ACS Photonics* **2016**, 3, 395–402. The RETOP freeware is available at the corresponding author group webpage.

# Supplementary Information for "Tailoring iridescent visual appearance with disordered resonant metasurfaces"

Adrian Agreda[1], Tong Wu[1], Adrian Hereu[2], Mona Treguer-Delapierre[2], Glenna L. Drisko[2], Kevin Vynck[3], Philippe Lalanne[1]*

[1]LP2N, CNRS, Institut d'Optique Graduate School, Univ. Bordeaux, F-33400 Talence, France
[2]CNRS, Univ. Bordeaux, Bordeaux INP, ICMCB, UMR 5026, F-33600 Pessac, France
[3]Institut Lumière Matière, CNRS, Université Claude Bernard Lyon 1, 69100 Villeurbanne, France
*Corresponding author: Philippe.Lalanne@institutoptique.fr

**Supplementary Note 1. Diffuse colors of iridescent metasurfaces for various spacer thicknesses**

This section demonstrates the color variability of iridescent metasurfaces for various SiO$_2$ film thicknesses.

Figure S1.1a shows the photographs of a set of metasurfaces that are recorded under directional broadband light illumination by a solar simulator (normal incidence) as the viewing angle varies. The pictures are taken for metasurfaces with $h = 0$ nm (without SiO$_2$ layer), $h = 105$ nm, $h = 410$ nm and $h = 710$ nm. For a simpler view, an arc-like angular scheme illustrates the color variations in Fig. S1.1b. The colors are displayed in the RGB color space and are obtained by averaging the RGB colors of the central part of the metasurface photographs in **a**. The SiO$_2$ layer clearly enlarges the available gamut of colors. In fact, it transforms a single baby-blue color ($h = 0$) varying little in shade with the viewing angle into a palette of non-trivial iridescences.

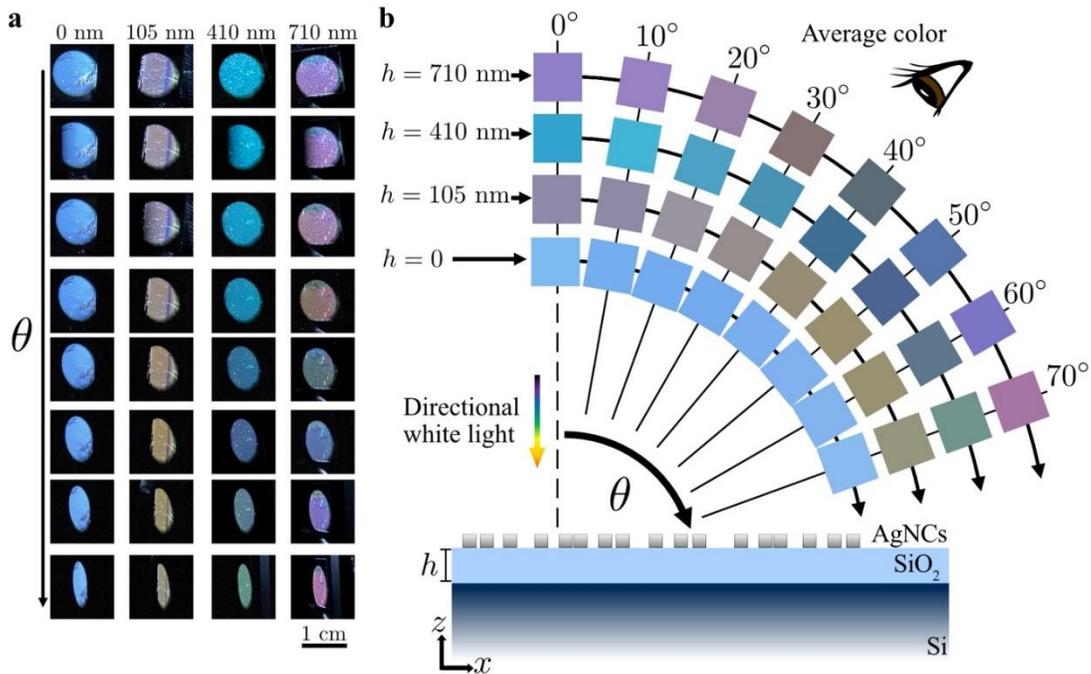

**Figure S1.1.** Diffuse iridescence for several SiO$_2$ layer thickness $h$. **a**, Real photographs of the samples recorded as the viewing angle $\theta$ is varied under illumination at normal incidence by a solar simulator.

**b**, Average colors extracted from the photographs in **a**. The color gamut significantly changes with the SiO$_2$ layer thickness $h$.

Figure S1.2 shows the spectrally resolved intensity maps of the density-normalized BRDFs of six metasurfaces illuminated at normal incidence. Due to the $\theta \rightarrow -\theta$ symmetry, only measurements for $\theta > 0$ are shown, $\theta = 0° - 70°$. Firstly, let us focus on the metasurface composed of randomly arranged Ag nanocubes directly deposited on the silicon substrate ($h = 0$). The map of Fig. S1.2a reveals a strong peak (see $\times 10^{-1}$ factor) concentrating most of the backscattered light around the specular direction at $\lambda < 430$ nm. As explained in the main text, the single peak comes from two localized plasmon resonances, one being dipolar and the other quadrupolar, which spectrally overlap owing to their electromagnetic hybridization with the substrate.

Likewise, the BRDF maps of a variety of stratified metasurfaces with Ag nanocubes on top of different SiO$_2$ layer thicknesses are shown in Fig. S1.2b-f. Cross-sectional SEM imaging and spectroscopic reflectometry studies provide estimates of the thicknesses in the range of $h \approx 105 - 710$ nm as indicated above the figures. The maps show distinctive resonance peaks that increase in number and vary in frequency as $h$ increases, tuning the color characteristics of the metasurfaces. Consistently with Fig. 3a, the presence of a thick silica layer turns two nearly overlapping resonances ($h = 0$) into several well-marked resonances at different wavelengths, which additionally switch on or off as the viewing angle changes. For $h \approx 320$ nm (Fig. S1.2d), for instance, two peaks at blue ($\lambda \approx 430$ nm) and red wavelengths ($\lambda > 600$ nm) give a magenta color at the zenith view, see Fig. S1.2b. As $\theta$ increases, the metasurface tends towards brownish colors due to the weakening of the blue peak. For $\theta = 60°$, much less light is scattered, and a grey color appears. For $h \approx 710$ nm, Fig. S1.2f, the metasurface exhibits richer blue, green, and red hues arising from the combination of three resonance peaks at different sets of wavelengths and viewing angles (see also Fig S5.1).

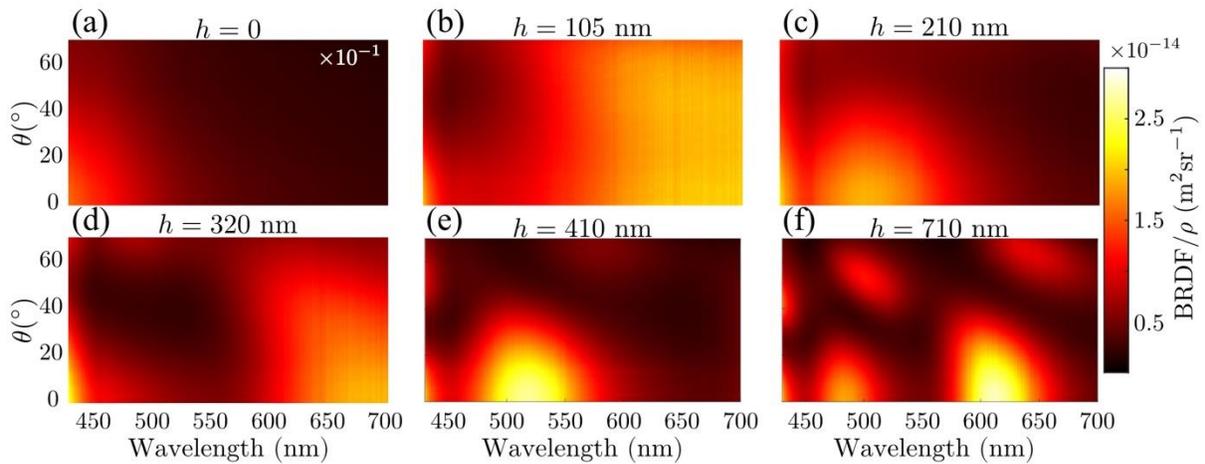

**Figure S1.2.** BRDF of the metasurfaces as a function of the wavelength and viewing polar angle $\theta$ under normal incidence ($\theta_i = 0$) with varying SiO$_2$ layer thicknesses, **(a)** $h = 0$ (the Ag nanocubes directly lay on the Si substrate), **(b)** $h \approx 105$ nm, **(c)** $h \approx 210$ nm, **(d)** $h \approx 320$ nm, **(e)** $h \approx 410$ nm and **(f)** $h \approx 710$ nm. Measurements are performed 10° above the plane of incidence. The BRDFs are normalized to the corresponding density of Ag nanocubes of each sample ($\rho \approx 0.2 - 8$ μm$^{-2}$).

## Supplementary Note 2. Marked difference between diffuse and classical thin-film iridescences

This section highlights that the color gamuts generated by the diffuse iridescence are completely different from the classical (specular) iridescence generated with the same $SiO_2$/Si substrate without nanocubes. It thus also highlights the key role played by the nanocube resonances in the diffuse iridescence of the metasurfaces and tempers the role of thin-film interference.

**Possible confusion between specular and diffuse iridescences.** The origin of the confusion is schematically illustrated in Fig. S2.1.

Figure S2.1a depicts a nanoparticle on a low-index thin layer film on a reflective substrate. The nanoparticle is illuminated at normal incidence (the direction is unimportant) and scatters light in multiple directions. Light directly scattered by the nanoparticle (ray 1 in Fig. S2.1a) and light subsequently reflected by the substrate (ray 2 in Fig. S2.1a) may lead to constructive or destructive interferences and different colors are observed as the viewing angle $\theta$ varies, irrespective of the incident angle.

In Fig. S2.1b, we consider classical thin-film iridescence. The monolayer of nanoparticles is replaced by a homogenized thin film (grey layer). Illuminated by a plane wave with an incident angle equal to the viewing angle ($\theta_{inc} = \theta$), the reflected rays 1 and 2 lead to interferences that are very similar to those in Fig. S2.1a. We thus anticipate, wrongly albeit logically, that the color gamuts generated by diffuse iridescence (a) for any incidence angle and specular reflection (b) are quasi-identical.

The intuitive prediction is completely erroneous: 1/ The color gamut of the diffuse iridescence phenomenon depends on the incidence angle, see Fig. 2b in the main text for instance. 2/ The color gamuts do not resemble each other, as will be experimentally evidenced in the next Sub-Section. The reason for the wrong prediction is that the diffuse iridescence mechanism depicted in Fig. S2.1a is oversimplified; it neglects the full set of nanoparticle resonances and their hybridization modes.

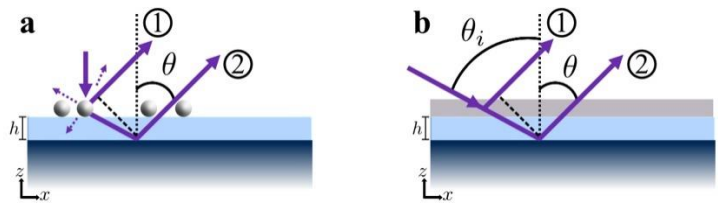

**Figure S2.1.** A naïve representation of the diffuse iridescence mechanism may lead to confuse diffuse and specular iridescence. **a**, Oversimplified diffuse iridescence mechanism. **b**, Specular iridescence.

**Diffuse and specular iridescences provide very different gamuts.** Figure S2.2 compares the specular and diffuse colors measured for the metasurfaces with a $SiO_2$ layer thickness $h = 320$ nm. We choose this thickness because it offers a large palette of colors (Fig. 2a).

The colored squares in Fig. S2.2a show the colors of the specular light captured by the smartphone camera at different angles of incidence/reflection ($\theta_i = \theta_r = 15° \rightarrow 60°$) in steps of $\Delta\theta_{i,r} = 5°$. Two quite similar sets of pictures are shown; the left set corresponds to measurements performed for the bare $SiO_2$/Si substrate (w/o AgNCs) and the right set corresponds to measurements performed for the AgNCs/$SiO_2$/Si metasurfaces. The recorded specular colors are completed by numerically computed and subsequently rendered colors of the specular reflected light from the bare $SiO_2$/Si substrate (colored arc).

The reflectance is computed with the 2×2 matrix-product approach [1] and the color rendering is performed with MATLAB. The two sets of experimental data and the numerical data are all consistent and indicate that the effect of the monolayer on the coherent (specular) light (modeled as a homogenized film in Fig. S2.1b) is negligible because of the low density of nanocubes.

In Fig. S2.2b, the diffuse colors of the metasurfaces are shown for $\theta_i = 0° \to 70°$ in steps of $\Delta\theta_i = 10°$ and scattering angles $\theta = -75° \to 80°$. The colors are captured with the smartphone camera every degree and averaged over an area of $16 \times 122$ pixels.

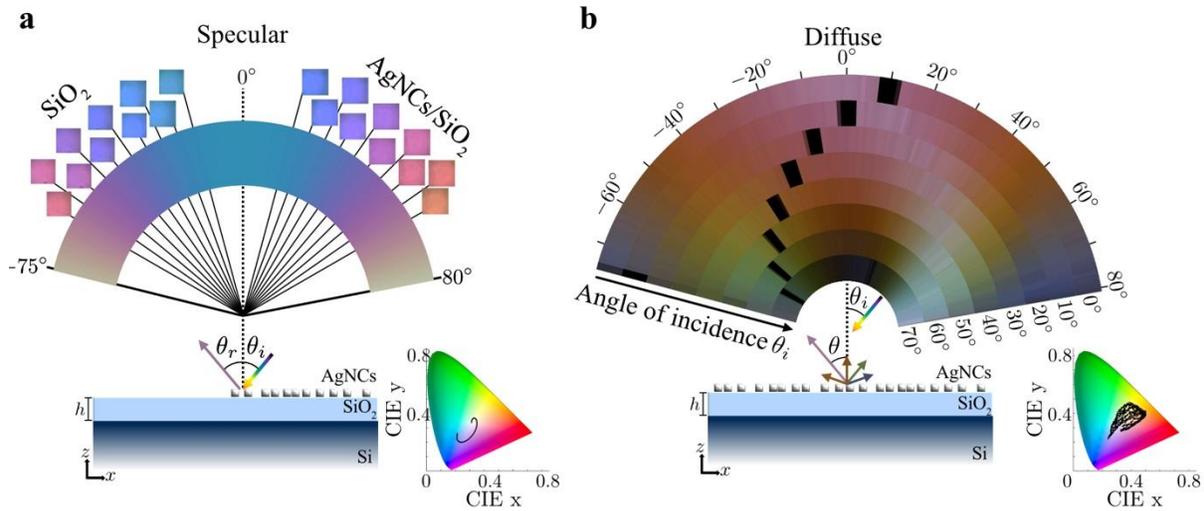

**Figure S2.2.** Specular and diffuse color gamut of the $h = 320$ nm metasurface. **a,** Specular light. Squares are experimental data collected for bare SiO₂/Si substrates (left) and AgNCs/SiO₂/Si metasurfaces (right) at different angles of incidence/reflection ($\theta_i = \theta_r = 15° \to 60°$) in steps of $\Delta\theta_{i,r} = 5°$. They correspond to the central part of the metasurfaces ($50 \times 50$ pixels). The arc corresponds to computed colors for bare SiO₂/Si substrates. **b,** Experimental diffuse colors of the AgNCs/SiO₂/Si metasurfaces as a function of the viewing and illumination angles. In **a-b,** the right insets show the CIE diagrams for each case.

The specular colors lie in the purple-blue-violet-pink-orange spectrum whereas the diffuse colors vary from blue to pink, passing through green and orange/yellow. As announced, the diffuse and specular colors are completely different. The right insets in Fig. S2.2 indicate the coordinates of the achieved colors in the CIE $xy$ chromaticity diagrams (Fig. S2.3 additionally shows the separated chromaticity diagrams of the diffuse light for each angle of incidence). The diagrams confirm the color differences and ratify the singularity and richness of the diffuse iridescence effect.

Despite the great and undeniable influence of the thin film, the diffuse iridescence mechanism reported here is quite different from the mechanism of ordinary thin-film iridescence. The individual resonances of the nanocube and their hybridization undoubtedly play an important role in the observed colors.

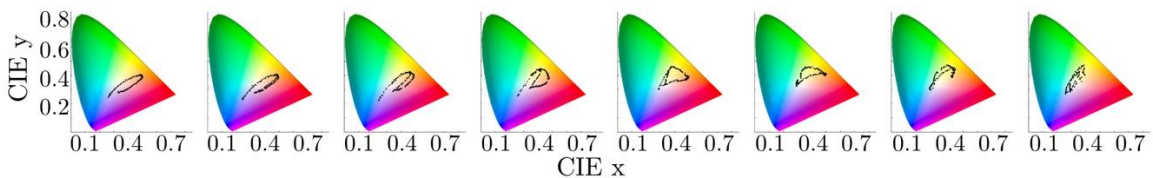

**Figure S2.3.** Chromaticity diagrams of the diffuse color panels shown in Fig. S2.2 from $\theta_i = 0°$ (leftmost panel) to 70° (rightmost panel).

## Supplementary Note 3. The nature of the plasmonic and hybridized modes

The individual nanocubes possess two dominant resonances, one with a dipolar (ED) character and the other one with a quadrupolar (EQ) character. On a reflective substrate, the ED resonances hybridize and give rise to Fabry-Perot (FP) resonances. In contrast, the EQ resonance does not hybridize and retains its plasmonic character. The hybridized (FP) and non-hybridized (plasmonic) resonances are different in nature and exhibit distinct properties, the most spectacular one being that the resonance frequency of the plasmonic mode cannot be tuned by varying the spacer thickness, whereas the FP resonance frequencies largely change as a function of thickness (Fig. 3a).

This Supplementary Section discusses other distinct properties, their markedly different excitation mechanisms or their far field radiations with different lobes, with the hope of better understanding how they can be used to control the iridescence. It also explains the origin of the splitting of the main BRDF peak as the incidence angle increases in Fig. S5.1a.

**Non-hybridized (plasmonic) resonances.** Assume an incident plane wave exciting the plasmonic resonance of an individual nanocube. Part of the light is directly scattered above in air, and the other part is scattered towards the substrate to be further backreflected (Fig. S3.1). Constructive interference occurs whenever

$$\frac{2\pi}{\lambda_{SP}} 2nh\cos(r) + \phi_r + \phi_{21} = 2m\pi, \quad (S3.1)$$

in which $m$ is a relative integer, $\lambda_{SP} \approx 432$ nm is the resonance frequency of the plasmonic mode, $n = 1.47$ is the refractive index of the SiO$_2$ thin film, and $r$ is the reflection angle in the thin film. The first term represents the difference between the orange and red optical paths in Fig. S3.1, $\phi_r \approx 180°$ is the phase caused by the reflection at the Si/SiO$_2$ interface, and $\phi_{21}$ is the phase difference between the light scattered towards the substrate (angle $r$) and that scattered in air (angle $\theta$, $\sin\theta = n\sin r$). $\phi_{21}$ can be computed with the RETOP software [2] from the far-field radiation of the EQ QNM obtained at the real frequency (see details in [3]) of the nanocube sitting on a SiO$_2$ substrate ($h = \infty$).

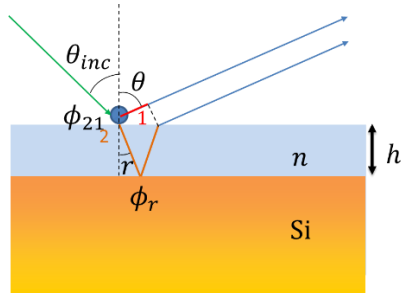

**Figure. S3.1.** The radiation of the plasmonic mode is determined by the interference between the light directly scattered above in air (ray 1) and that scattered towards the substrate and subsequently reflected (ray 2). $\phi_{21}$ is the phase caused by the scattering of the nanoparticle; $\phi_r \approx 180°$ is the phase caused by the reflection of the Si/SiO$_2$ interface. The mechanism is independent of the incident beam (polarization or $\theta_{inc}$) as $\phi_{21}$ is intrinsic to the QNM.

The radiation of the EQ mode is maximal around $r \approx 43°$, see Fig. 3d, but this angle corresponds to the total internal reflection. Therefore, the condition will be satisfied for smaller $r$'s. From Eq. (S3.1), we readily find the constructive interference directions:

$$\sin(\theta) = n \sin\left(\text{acos}\left(\frac{[2\pi m + \phi_r + \phi_{21}]\lambda_{SP}}{4\pi n h}\right)\right). \tag{S3.2}$$

The variation of $\phi_{21}$ with $\theta$ can be neglected and $\phi_{21} + \phi_r$ can be approximately taken as 36° according to our calculation. The solid curves in Fig. S3.2 shows $\theta$ as a function of the spacer thickness $h$ for $m = 1, 2, …4$. To test our simple model, we have superimposed the far field radiation diagrams of the plasmonic modes for various $h's$. A qualitative agreement is obtained between the predictions of Eq. (S3.2) and the directions of the dominant lobes.

We emphasize that, in contrast with the hybridized modes (see below), the direction or polarization of the incident light does not play any role into the QNM far-field radiation diagram.

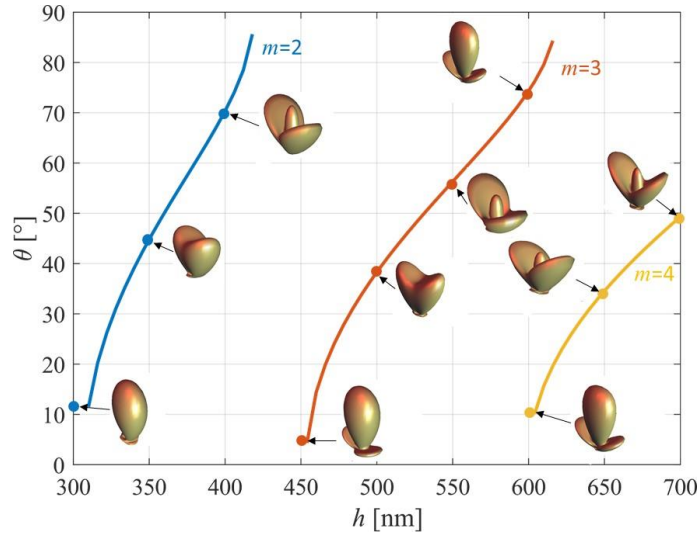

**Figure S3.2.** Test of Eq. (S3.1) for $m = 1, 2, 3, 4$. The scattering diagrams of the normalized QNMs are computed at the real frequency $\lambda_{SP} \approx 432$ nm [3].

**Hybridized (FP) resonances.** To simplify the analysis, we assume that the Si substrate behaves as a perfect electric conductor. The hybrid modes then consist of a pair of symmetric or antisymmetric coupled QNMs. For a polarization parallel to the interface, the mirror image is antisymmetric (Fig. S3.3).

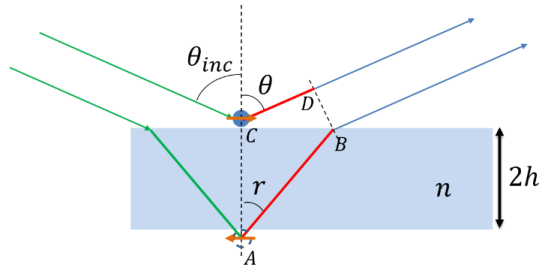

**Figure S3.3.** The radiation of the hybrid mode is determined by the interference between the light scattered by the nanoparticle dipole and that scattered by the image dipole. In a perfect electric conductor approximation, the FP mode is anti-symmetric (orange arrows) owing to the tangential ED

nature of the hybridized QNM. The hybrid mode can be efficiently excited by a plane wave with an incident angle $\theta_{inc} = \theta$.

The scattering angle $\theta$ in air for which constructive interference occurs are obtained when the path difference $nAB - CD = 2hn \cos r$ is equal to $(\frac{1}{2} + m)\lambda_{FP}$

$$\sin(\theta) = n \sin\left(\mathrm{acos}\left(\frac{2\pi\left(m+\frac{1}{2}\right)\lambda_{FP}}{4\pi n h}\right)\right). \tag{S3.3}$$

Figure S3.4 shows the predicted scattering angle $\theta$ as the spacer thickness is varied for the third Fabry-Perot branch in Fig. 3a (the one for which 3 radiation diagrams are displayed). Again, the formula quantitatively predicts the directions of the main lobes of the far-field radiation diagrams.

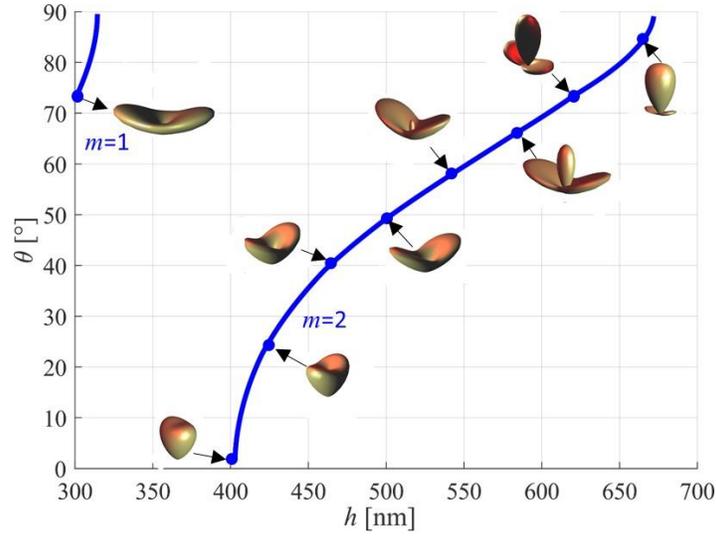

**Figure S3.4.** Test of Eq. (S3.3) for $m = 1,2$. The scattering diagrams of the normalized QNMs are computed at the real frequency $\lambda_{FP}$ of the QNMs of the third FP branch in Fig. 3a (main text).

**Comparison of Eqs. (S3.2) and (S3.3).** Equations (S3.3) and (S3.2) are quite similar. However, this apparent similarity hides a few fundamental differences. First, unlike the plasmonic mode, the resonance wavelength $\lambda_{FP}$ of the hybridized QNM now varies with $h$. Second, the phase factors are different. In the non-hybridized case, the phase factor $\phi_{21}$ is determined from the non-trivial scattering properties of individual nanocubes laying on a SiO$_2$ substrate; $\phi_{21}$ can take an arbitrary value a priori. In contrast, for hybridized modes, since the nanoparticle dipole and its mirror image are phase matched, the phase $1/2(2\pi)$ in Eq. S3.3 is locked at $180°$ (with the perfect electric conductor approximation) and is independent of the nanoparticle size and shape.

Lastly, there is another important difference concerning the excitation condition by an incident plane wave. For the FP QNM, an effective excitation requires the driving fields at the locations of the nanoparticle and its mirror image to be $\pi$ out-of-phase (according to the perfect electric conductor model). In sharp contrast with the plasmonic resonance, such a condition is satisfied for an incident plane wave with a frequency equal to the resonance frequency and with an incident angle equal to the scattering angle, $\theta_{inc} = \theta$, see Fig. S3.3. The scattered and incident directions are entangled by hybridization. This property, which is analogous to a phase matching condition, will be used in the Supplementary Note 5 to explain why the long-wavelength FP peak splits in two peaks centered at $\theta = \pm\theta_{inc}$ as the angle of incidence $\theta_{inc}$ increases in Fig. 3a-c (main text) or in Fig. S5.1a.

## Supplementary Note 4. Analysis of the hue difference.

Figure S4.1 is a polar representation indicating the hue value at normal view and at the viewing angle with the farthest hue value. This is done for every angle of incidence (hand fan rings of Figure 2). The points are colored with the sRGB corresponding to that position. The HSV (which stands for hue, saturation, value) values come from the sRGB values extracted from the photographs and converted by color matrix transformation. We consider the hue difference to be underestimated in some cases since photographs are limited to the sRGB space. However, the figure clearly confirms the large hue differences for the bicolored metasurface ($h = 640$ nm) with an almost constant value at normal view (green hues). Hue differences at the largest angle of incidence ($\theta_i > 50°$) are probably biased by the low brightness of the metasurface.

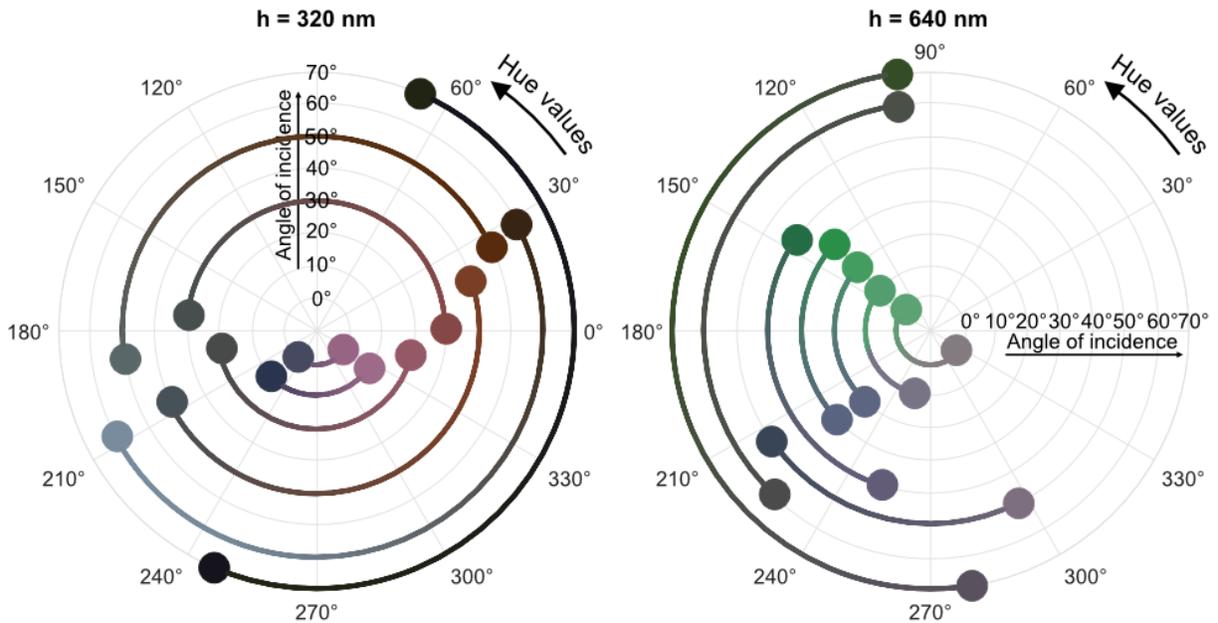

**Figure S4.1.** Polar representation of the hue variations for $h = 320$ nm and $h = 640$ nm iridescent metasurfaces for every angle of incidence shown in Fig. 2b. The color of the points is given by the sRGB value captured by the camera. The HSV values come from the sRGB values extracted from the photographs and converted by color matrix transformation.

## Supplementary Note 5. Appearance of curved metasurfaces.

In the main text, we focus on the appearance of flat metasurfaces, and it is tempting to consider curved metasurfaces. For instance, non-flat objects could be covered with metasurfaces in the future with plastic film applied to a metallic layer. Owing to the curvature, the normal to the metasurface is no longer fixed, and, depending on the local normal, we expect to observe a non-uniform hue forming a patchwork dominantly composed of violet and green zones. In Fig. S5.1, we compare the hues observed with our flat two-color metasurface (AgNCs/SiO₂/Si, $h = 640$ nm) under sun-light illumination with a series of rendered images recently reported in [4] by some of the authors of the present work. The renders show a curved object, a *speedshape*, covered with a metasurface that is quite similar to the present one: same materials, spacer thickness and nanoparticle volume. Despite some slight differences concerning the illuminant sources and particle shapes as well as polydispersity that is neglected in the computation, the observations of our flat samples (Fig. S5.1a) agree well with the synthetic images (Fig. S5.1b).

This retrospectively confirms the accuracy of the numerical tool developed in [4] to visualize the appearance of complex metasurfaces at the macroscale. We thus expect that a curved object covered with the two-color metasurface would exhibit a patchwork of green and violet zones, as shown in (b), which gently deform and glide on the curved surface as the viewing direction is varied. Watch the leftmost Supplementary Video 1 in [4] to see this remarkable visual effect for the speedshape.

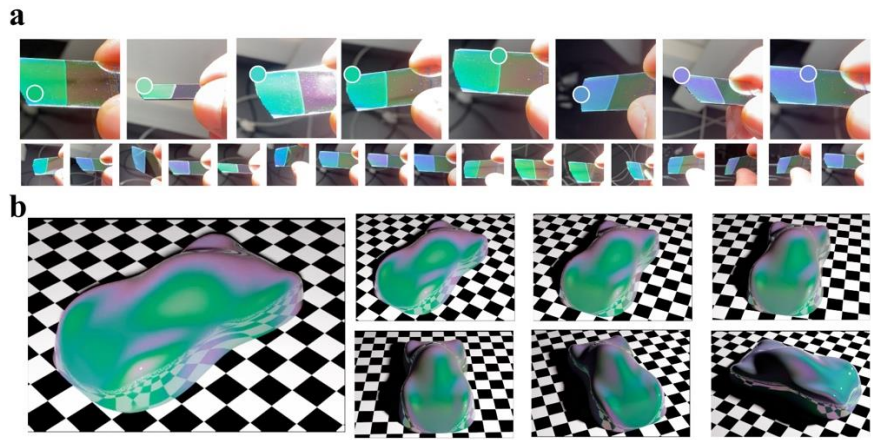

**Figure S5.1.** Fabricated vs realistically rendered diffuse iridescent metasurfaces. **a**, Series of 25 photographs of the two-color metasurface under sunlight illumination during a clear day. The sample is shown at several random orientations to vary the angles of incidence and view (see Supplementary Video SV1 for a better and continuous visualization). **b**, Rendered images of a speedshape covered by Ag nanospheres (with the same volume as the Ag nanocubes) on a SiO₂/Si substrate for $h = 600$ nm illuminated by a directional sun-like light (illuminant E) at 7 viewing directions, see [4] for details.

The comparison made here is purely qualitative and several aspects influencing the visual perception of the fabricated metasurface are left aside. The complexity of the illumination environment and the detection system settings are not fully considered at this point. A comprehensive discussion about the visual appearance differences between the fabricated and realistically rendered metasurfaces is beyond the scope of this work.

## Supplementary Note 6. Main properties of the BRDF maps

This section shows additional BRDFs of the two-color metasurface for several angles of incidence, $\theta_i = 0°, 10° \ldots 60°$ (Fig. S6.1a) and explains the physical origin of their three prominent features: asymmetry, blue-shift and splitting of the peaks as $\theta_i$ increases.

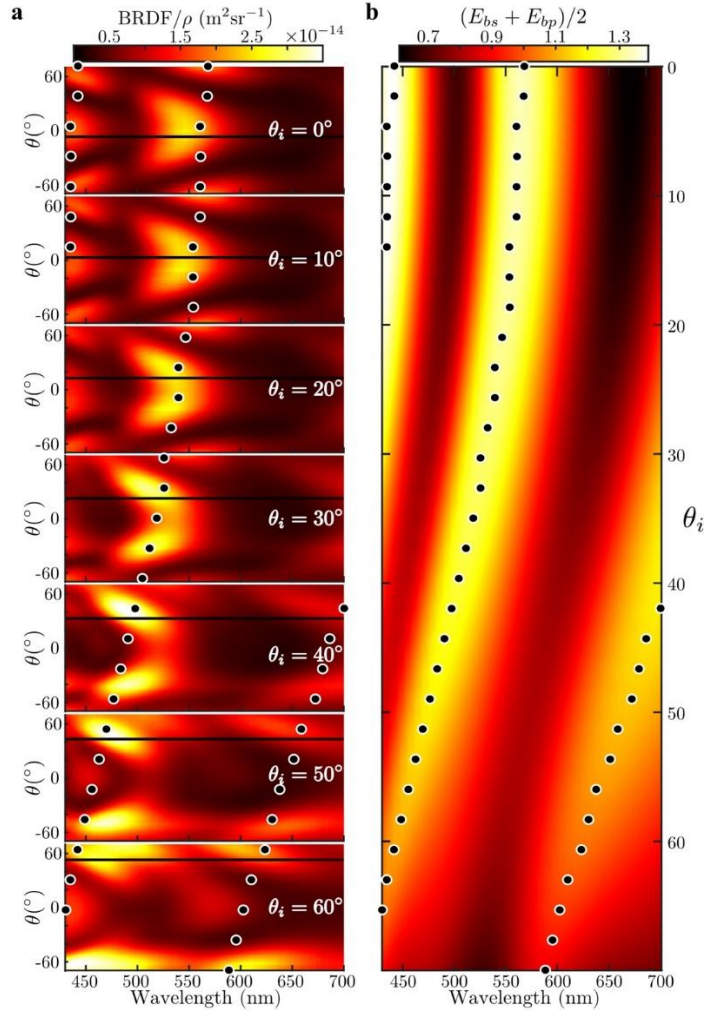

**Figure S6.1.** Influence of the incident angle on the BRDF maps of the two-color metasurface. **a**, Measured normalized BRDF maps as a function of wavelength and scattering angle $\theta$ for $\theta_i = 0°, 10° \ldots 60°$. The BRDFs are measured at $\alpha = 10°$ above the plane of incidence (See Fig. S6.1b). The black dots correspond to the intensity peaks of the background field in **b**. **b**, Computed intensity of the background field at the nanocube center for $h = 640$ nm as a function of the incident angle and wavelength. The peak maxima for unpolarized light are marked with black dots.

**Asymmetry.** The BRDFs at large $\theta_i$'s (typically $\theta_i \geq 50°$) no longer satisfy the $\theta \to -\theta$ symmetry, especially in the spectral interval from 500 to 600 nm. Helped by the QNM model, we attribute the asymmetry in the excitation to a vertically polarized ED mode of the Ag nanocube, which can be strongly excited only by TM-polarized waves and large incident angles. The field scattered by this QNM interferes with those of the EQ and FP modes. Since constructive (resp. destructive) interferences occur for $\theta > 0$ (resp. $\theta < 0$) directions, the $\theta \to -\theta$ symmetry is broken.

**Blue shift.** An important trend in Fig. S6.1a is the systematic blue shift of all the peaks as $\theta_i$ increases. The trend, which is indicated by the black dots, is easily understood with the QNM model. The later predicts that the spectral peaks of the BRDF arise not only from the resonances themselves, but also from the frequencies for which they are effectively excited. As shown in the main text, the QNM excitation strength $\alpha_m(\omega) = L_m(\omega)I_m(\omega)$ is proportional to $\mathbf{E}_b(\mathbf{r}, \omega)$ through the overlap integral $I(\omega)$. Thus, the BRDF peaks are preferentially observed at frequencies for which $|\mathbf{E}_b(\mathbf{r},\omega)|$ is maximum. To verify our intuition, we compute $|\mathbf{E}_b(\mathbf{r},\omega)|$ for unpolarized light with the $2\times 2$ matrix-product approach [1]. The computed data are displayed in Fig. S6.1b as a function of $\lambda$ and $\theta_i$ and the peak maxima are marked with black dots. The blue shift of the BRDF peaks is quantitatively recovered with this simple computation, the small mismatch being attributed to the Lorentzian factor $L_m(\omega)$ of the excitation coefficient.

**Peak splitting.** Yet another important feature of the BRDF maps is the evolution of the dominant FP peak (the peak at 540 nm at $\theta_i = 0°$) as $\theta_i$ increases. The peak splits in two peaks symmetrically centered around $\theta = 0$ and the separation between the peaks progressively increases as $\theta_i$ increases. It is noticeable that the angular positions of the split peaks coincide with the horizontal black strips that correspond to detection angles for which the incident light is obscured by the detector support, implying that the split peaks are observed for $\pm\theta \approx \theta_i$. This can be understood by considering the nature of the hybridized QNMs. As explained at the end of the Supplementary Note 3, the scattered ($\theta$) and incident ($\theta_{inc}$) directions are entangled by hybridization, implying that FP modes are efficiently excited whenever the direction of the incident plane wave coincides with the preferential scattering direction of the hybridized QNM. This exactly corresponds to $\pm\theta \approx \theta_i$.

## Supplementary Note 7. Goniometric setup and reliability of the measurements.

All the BRDF measurements are performed with the same experimental apparatus (See Fig. S7.1). The reflected light is collected at $\alpha = 10°$ above the plane of incidence with a spectrometer connectorized with a 1 mm diameter fiber. The detection/viewing angle, $\theta$, of the goniometer is varied from $-70°$ to $70°$. A spectrum is then recorded every $\Delta\theta = 1°$. The size of the incident laser spot is around 1 cm. The light is filtered spatially to ensure a spectrally homogeneous beam. Neutral density filters control the laser intensity to avoid saturation of the spectrometer. The incident light is measured by placing the detector facing the focused laser beam. The set-up also allows a camera to be mounted, instead of a fiber detector, to visually capture the appearance of the sample.

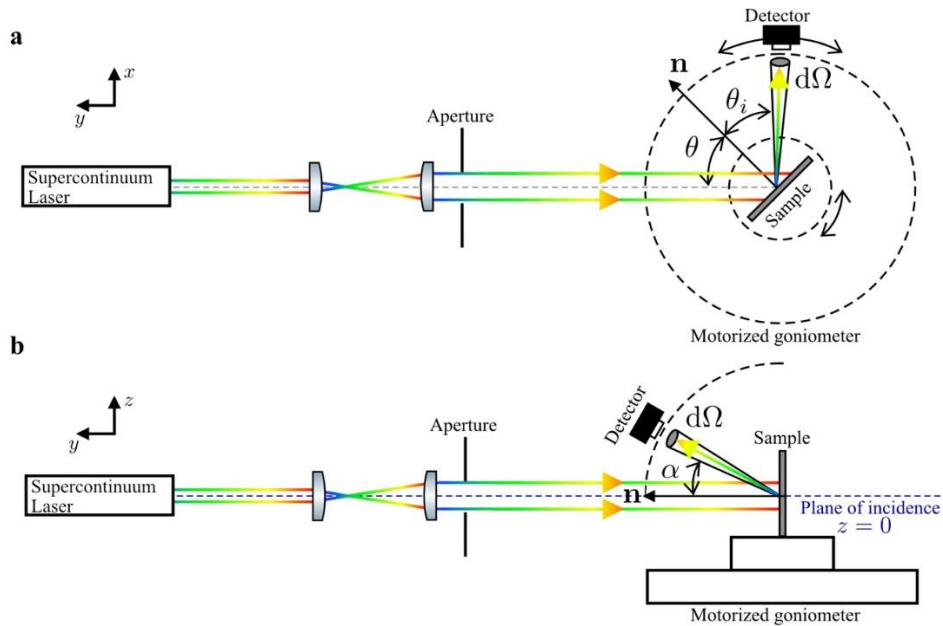

**Figure S7.1.** BRDF goniometer set-up in **a**, top and **b**, profile view. A super continuum laser is expanded and size controlled by a circular aperture to homogeneously illuminate the sample. The angle of incidence is controlled by a rotational motor and the reflected light is collected by an optical fiber mounted in another rotational motor and connected to a spectrometer.

BRDF measurements on a standard reference sample with a known reflectance have confirmed the accuracy of the measurements. A Spectralon (Labsphere SRS-99-010) is placed at the sample holder and illuminated at normal incidence with a supercontinuum laser beam (Fig. S7.1). The measured intensity is integrated over the wavelengths (left figure). In total, five runs of measurements are performed. The missing point at $\theta \approx 10°$ comes from the holder arm of the detector which blocks the laser light. The measured values (Fig. S7.2) approximate very well the expected theoretical value of $1/\pi$ for a perfect Lambertian reflector (horizontal dashed line) with an expected diminished intensity for larger angles. Furthermore, the results are compared with those obtained with another Spectralon (also from Labsphere) in [5] for $\lambda = 680$ nm (right panel). The discrepancy between the values reported in [4] and our measurements does not exceed 10%.

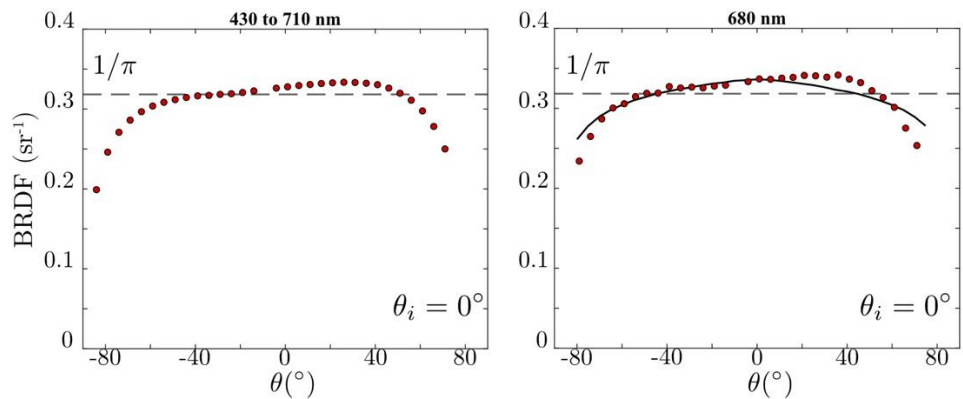

**Figure S7.2.** Spectralon BRDF for the visible spectrum $\lambda = 410 - 710$ nm (Left) and $\lambda = 680$ nm (Right). The horizontal dashed line at $\mathrm{BRDF} = 1/\pi$ indicates the theoretical expected value. For comparison, in the right panel, our measurements are compared with those obtained in [5] (solid black curve) for the same wavelength ($\lambda = 680$ nm).

## Supplementary References

1. Yeh, P., *Optical waves in layered media*, J. Wiley and Sons eds., New York, 1988.
2. Yang, J., Hugonin, J. P. & Lalanne, P. Near-to-far field transformations for radiative and guided waves. ACS Photon. **3**, 395–402 (2016). The associated Matlab software RETOP can be downloaded from the group webpage of the last author and is available on Zenodo.
3. Wu, T., Arrivault, D., Yan, W., Lalanne, P. Modal analysis of electromagnetic resonators: User guide for the MAN program, Comput. Phys. Commun. **284**, 108627 (2023).
4. Vynck, K., Pacanowski, R., Agreda, A., Dufay, A., Granier, X. & Lalanne, P. The visual appearances of disordered optical metasurfaces. Nat. Mater. **21**, 1035-1041 (2022).
5. Bhandari, A., Hamre, B., Frette, Ø., Zhao, L., Stamnes, J. J., & Kildemo, M. Bidirectional reflectance distribution function of Spectralon white reflectance standard illuminated by incoherent unpolarized and plane-polarized light. Appl. Opt. **50**, 2431-2442 (2011).